\documentclass{JHEP3}    
\usepackage{epsfig}
\usepackage{subfigure}
\usepackage{psfrag}
\usepackage{graphicx}

\newcommand{\eref}[1]{(\ref{#1})}

\newcommand{\fref}[1]{Figure~\ref{#1}}
\newcommand{\cref}[1]{Chapter~\ref{#1}}
\newcommand{\beq}{\begin{equation}}
\newcommand{\eeq}{\end{equation}}
\newcommand{\ba}{\begin{array}}
\newcommand{\ea}{\end{array}}
\newcommand{\bcenter}{\begin{center}}
\newcommand{\ecenter}{\end{center}}

\def\IB{\relax\hbox{$\inbar\kern-.3em{\rm B}$}}
\def\IC{\relax\hbox{$\inbar\kern-.3em{\rm C}$}}
\def\ID{\relax\hbox{$\inbar\kern-.3em{\rm D}$}}
\def\IE{\relax\hbox{$\inbar\kern-.3em{\rm E}$}}
\def\IF{\relax\hbox{$\inbar\kern-.3em{\rm F}$}}
\def\IG{\relax\hbox{$\inbar\kern-.3em{\rm G}$}}
\def\IGa{\relax\hbox{${\rm I}\kern-.18em\Gamma$}}
\def\IH{\relax{\rm I\kern-.18em H}}
\def\IK{\relax{\rm I\kern-.18em K}}
\def\IL{\relax{\rm I\kern-.18em L}}
\def\IP{\relax{\rm I\kern-.18em P}}
\def\IR{\relax{\rm I\kern-.18em R}}
\def\IZ{\relax\ifmmode\mathchoice
{\hbox{\cmss Z\kern-.4em Z}}{\hbox{\cmss Z\kern-.4em Z}}
{\lower.9pt\hbox{\cmsss Z\kern-.4em Z}}
{\lower1.2pt\hbox{\cmsss Z\kern-.4em Z}}\else{\cmss Z\kern-.4em Z}\fi}
\def\II{\relax{\rm I\kern-.18em I}}

\def\sCC{{\kern 0.27em\vrule height1.45ex width0.03em depth0em
          \kern-0.30em\rm C}}
\def\C{{\mathchoice
  {\sCC}
  {\sCC}
  {\kern 0.225em \vrule height1.05ex width0.025em depth0em \kern-0.25em \rm C}
  {\kern 0.180em \vrule height0.78ex width0.02em depth0em \kern-0.2em \rm C}
        }}
\def\sHH{{\rm I\kern-.16em{}H}}
\def\H{{\mathchoice
  {\sHH}
  {\sHH}
  {\rm I\kern-.13em{}H}
  {\rm I\kern-.13em{}H} }}
\def\sNN{{\rm I\kern-.16em{}N}}
\def\N{{\mathchoice
  {\sNN}
  {\sNN}
  {\rm I\kern-.12em{}N}
  {\rm I\kern-.10em{}N} }}
\def\sPP{{\rm I\kern-.16em{}P}}
\def\P{{\mathchoice
  {\sPP}
  {\sPP}
  {\rm I\kern-.12em{}P}
  {\rm I\kern-.10em{}P} }}
\def\sQQ{{\kern 0.27em \vrule height1.45ex width0.03em depth0em
          \kern-0.30em \rm Q}}
\def\Q{{\mathchoice
        {\sQQ}
        {\sQQ}
  {\kern 0.225em \vrule height1.05ex width0.025em depth0em \kern-0.25em \rm Q}
  {\kern 0.180em \vrule height0.78ex width0.020em depth0em \kern-0.20em \rm Q}
        }}
\def\sRR{{\rm I\kern-0.16em{}R}}
\def\R{{\mathchoice
  {\sRR}
  {\sRR}
  {\rm I\kern-0.12em{}R}
  {\rm I\kern-0.10em{}R} }}
\def\sZZ{{\rm Z\kern-0.32em{}Z}}
\def\Z{{\mathchoice
  {\sZZ}
  {\sZZ} 
  {\rm Z\kern-0.3em{}Z}     
  {\rm Z\kern-0.25em{}Z} }}  
\def\ZZZ{{\rm Z\kern-0.24em{}Z}}
\def\sII{{\rm I\kern-0.16em{}I}}
\def\I{{\mathchoice
  {\sII}
  {\sII}
  {\rm I\kern-0.12em{}I}
  {\rm I\kern-0.10em{}I} }}

\def\Tr{{\rm Tr}}

\def\vol{{\rm vol}}

\def\inbar{\,\vrule height1.5ex width.4pt depth0pt}
\font\cmss=cmss10 \font\cmsss=cmss10 at 7pt

\def\smiley{\hbox{\large$\bigcirc$\hspace{-0.80em}\raise.2ex
\hbox{$\cdot\cdot$}\kern-.61em\lower.2ex\hbox{\scriptsize$\smile$}}\ }
\def\frowny{\hbox{\large$\bigcirc$\hspace{-0.80em}\raise.2ex
\hbox{$\cdot\cdot$}\kern-.635em\lower.2ex\hbox{\scriptsize$\frown$}}\ }

\def\I{{\rlap{1} \hskip 1.6pt \hbox{1}}}

\makeatletter
\let\hangafter\@hangfrom
\makeatother

\newcommand{\drawsquare}[2]{\hbox{%
\rule{#2pt}{#1pt}\hskip-#2pt
\rule{#1pt}{#2pt}\hskip-#1pt
\rule[#1pt]{#1pt}{#2pt}}\rule[#1pt]{#2pt}{#2pt}\hskip-#2pt
\rule{#2pt}{#1pt}}

\newcommand{\fund}{\raisebox{-.5pt}{\drawsquare{6.5}{0.4}}}
\newcommand{\antifund}{\overline{\fund}}

\newcommand{\be}{\begin{equation}}
\newcommand{\ee}{\end{equation}}
\newcommand{\bea}{\begin{eqnarray}}
\newcommand{\eea}{\end{eqnarray}}
\newcommand{\bean}{\begin{eqnarray*}}
\newcommand{\eean}{\end{eqnarray*}}
\newcommand{\IS}{\bf S}
\newcommand{\beqa}{\begin{eqnarray}}
\newcommand{\eeqa}{\end{eqnarray}}
\newcommand{\id}{\bf 1}

\def\tr{{\rm tr \,}}
\def\Tr{{\rm Tr \,}}
\def\NN{{\cal N
}}

\def\CM {{\cal M}}

\preprint{CERN-PH-TH/2006-060 \\ IFT-UAM/CSIC-06-17 \\ PUPT-2195}

\title{Dynamical SUSY Breaking at Meta-Stable Minima from 
D-branes at Obstructed Geometries}

\author{Sebasti\'an Franco${}^1$ and Angel M .Uranga ${}^2$

\\
~\\
${}^1$Joseph Henry Laboratories, Princeton University,
Princeton, NJ  08544, USA 
\footnote{Research supported by National Science Foundation Grant No. 
PHY-0243680}
\\ 

${}^2$
PH-TH Division, CERN, CH-1211 Geneva 23, Switzerland \\
and Instituto de F\'{\i}sica Te\'orica, C-XVI, UAM, 28049 Madrid, Spain
\footnote{Research supported by CICYT, Spain, under project 
FPA-2003-02877, and the RTN networks MRTN-CT-2004-503369 `The Quest for 
Unification: Theory confronts Experiment' and MRTN-CT-2004-005104 
`Constituents, Fundamental Forces and Symmetries of the Universe'.}

\email{sfranco@feynman.princeton.edu, angel.uranga@cern.ch} \\
}

\abstract{
We study the existence of long-lived meta-stable supersymmetry breaking 
vacua in gauge theories with massless quarks, upon the addition of 
extra massive flavors. A simple realization is provided by
a modified version of SQCD with $N_{f,0}<N_c$ massless flavors, $N_{f,1}$ massive flavors
and additional singlet chiral fields. This theory has local meta-stable 
minima separated from a runaway behavior at infinity by a potential 
barrier. We find further examples of such meta-stable minima in 
flavored versions of quiver gauge theories on fractional 
branes at singularities with obstructed complex deformations, and study 
the case of the $dP_1$ theory in detail. Finally, we provide an explicit 
String Theory construction of such theories. The additional flavors 
arise from D7-branes on non-compact 4-cycles of the singularity, for which 
we find a new efficient description using dimer techniques.
}
\begin{document}

\section{Introduction}

The realization of supersymmetric gauge field theories on the world-volume 
of D-brane configurations in String Theory has proved to be an extremely 
insightful tool in the study of non-trivial gauge dynamics. In the context 
of $\mathcal{N}=1$ supersymmetric gauge field theories, an interesting class of 
models is obtained by considering systems of D3-branes at Calabi-Yau 
singularities, possibly in the presence of fractional branes. 
The resulting quiver gauge theories lead, in the absence of fractional
branes, to a tractable class of 4d strongly coupled conformal field theories, 
which extend the AdS/CFT correspondence \cite{Maldacena:1997re,Gubser:1998bc,Witten:1998qj}
to theories with reduced (super)symmetry \cite{Kachru:1998ys,Klebanov:1998hh,Morrison:1998cs} and
enable non-trivial precision tests of the correspondence (see for
instance \cite{Bertolini:2004xf,Benvenuti:2004dy})
\footnote{In the past few years there has been tremendous progress
in our understanding of AdS/CFT dual pairs. In addition to the papers
mentioned in the introduction, some of the works
that have been crucial for these developments are \cite{Gauntlett:2004yd}-\cite{Martelli:2006yb}.}.
Quiver gauge theories constructed with both D3-branes and 
fractional branes are not conformal and their RG flow involves 
cascades of Seiberg dualities \cite{Klebanov:2000hb,Franco:2003ja,Franco:2004jz,Franco:2005fd,Herzog:2004tr}. 
Finally, quiver gauge theories on fractional branes lead to non-conformal theories, with non-trivial 
strong dynamics effects, like confinement, or appearance of 
non-perturbative superpotentials 
\cite{Klebanov:2000hb,Franco:2005fd,Berenstein:2005xa,Franco:2005zu,Bertolini:2005di}.

One of the most interesting strong dynamics effects in $\mathcal{N}=1$ 
supersymmetric gauge field theories, both from the theoretical and the 
phenomenological viewpoints, is Dynamical Supersymmetry Breaking. It is 
thus natural to ask whether it can be realized on the world-volume of 
configurations of D-branes. Strictly speaking, dynamical supersymmetry 
breaking requires the removal of the classical supersymmetric vacuum, and 
the appearance of a global non-supersymmetric minimum of the potential. 
D-brane configurations realizing this phenomenon have not been found yet. For 
instance, gauge theories on certain fractional branes at geometries 
without complex deformations have been shown to develop non-perturbative 
superpotentials which remove the supersymmetric vacuum 
\cite{Berenstein:2005xa,Franco:2005zu,Bertolini:2005di}. However, the 
scalar 
potential of these theories, at least in the large field region where the 
K\"ahler potential can be trusted, leads to a runaway to infinity
\cite{Franco:2005zu} (see also \cite{Intriligator:2005aw,Brini:2006ej}).

An interesting alternative proposal is that dynamical supersymmetry 
breaking occurs at local meta-stable minima, separated from 
supersymmetric vacua by a large potential barrier. This idea, 
which first appeared in phenomenological model building (see e.g. 
\cite{Dimopoulos:1997ww}) has been realized in 
\cite{Intriligator:2006dd} in a strikingly simple system. The 
authors show that the introduction of massive flavors to $SU(N)$ SYM, with 
masses much smaller that the dynamical scale of the gauge sector, leads to 
the appearance of such a local meta-stable minimum, separated from the $N$ 
supersymmetric minima by a potential barrier. Furthermore, the non-supersymmetric
minimum can be made parametrically long-lived. We refer to this theory as 
the ISS model.

It is a natural question whether the introduction of extra massive 
flavors in more involved gauge theories also leads to such local 
meta-stable minima. In particular it would be interesting to explore this 
question for gauge theories realized on D-branes. These are the 
questions we address in the present paper. We find interesting 
generalizations of the ISS proposal, and find that the introduction of 
extra massive flavors leads to the appearance of local non-supersymmetric 
minima in diverse gauge theories with massless flavors. These theories  
include a simple extended version of $SU(N)$ SQCD, and the gauge 
theory on fractional branes on the complex cone over $dP_1$ (and related 
examples). Moreover, we argue 
that such local minima are likely to appear in quiver gauge theories of 
fractional branes in obstructed geometries (the so-called DSB fractional 
branes \cite{Franco:2005zu}).

The study of such generalization requires the development of new results 
in several directions, which are of interest in their own right, and  
which we provide in the present paper. The main results are as follows:

\medskip

$\bullet$ Since quiver gauge theories contain 
massless bi-fundamentals, it is first necessary to consider the 
generalization of the ISS proposal to theories with massless flavors. 
Hence, we study the introduction of additional massive flavors to 
SQCD with massless flavors in detail. We show that this theory does not have local 
meta-stable minima at one-loop, in contrast with the ISS case. 

\medskip

$\bullet$ We consider a simple extension of SQCD with massless flavors, 
by introducing extra fields with cubic coupling to the flavors. We study 
the introduction of additional massive flavors in this theory in detail, 
and show the appearance of local meta-stable minima. Interestingly this 
extended theory, which naturally generalizes the ISS proposal to theories 
with massless flavors, is tantalizingly similar to the quiver gauge 
theories on fractional branes at obstructed geometries. It hence provides an 
excellent toy model of the behavior for the latter.

\medskip

$\bullet$ We then consider the quiver gauge theory on fractional 
branes at the simplest example of an obstructed geometry, namely the 
complex cone over $dP_1$ (equivalent to the real cone over $Y^{2,1}$). We 
carry out the gauge theory analysis of this $dP_1$ theory upon the 
addition of extra massive flavors, and show the appearance of a 
non-trivial local minimum separated by a potential barrier from the 
supersymmetric minimum at infinity (equivalently, from the runaway 
behavior at large fields). The structure of fields and couplings, key to 
the existence of this minimum, is a general feature of gauge theories on 
fractional branes at obstructed geometries, strongly suggesting a 
generalization to this full class.

\medskip

$\bullet$ Finally, an explicit construction of this gauge theory in String 
Theory requires a D-brane realization of the incorporation of 
massive flavors. This is naturally achieved by the introduction of 
D7-branes in the configurations, which however has not been discussed in 
the literature for the case of general toric singularities. We carry out 
this analysis and provide new tools to introduce such D7-branes and 
easily determine the structure of new flavors from D3-D7 open strings, and 
their interactions. The flavor mass terms receive a natural interpretation 
in terms of vevs for higher dimensional scalars in the D7-D7 sector, which trigger 
a geometrical process that recombines several D7-branes, separating them 
from the D3-branes at the singularity.

\medskip

The outcome is that gauge theories on fractional branes at 
obstructed geometries provide a natural generalization of the ISS proposal 
to quiver gauge theories. Although the computational difficulties allow us 
to establish this result only in particular examples, we find convincing 
evidence that the picture is far more general. We expect that future work 
in this direction confirms this expectation.

The above results are discussed in different sections.
Some of them are presented as appendices to simplify the reading. The 
paper is organized as follows.

In Section \ref{sqcd} we study SQCD-like theories. Section 
\ref{puresqcd} reviews the ISS model for SYM with extra massive flavors. 
Section \ref{flavoursqcd} studies the introduction of massive flavors in 
SQCD with massless flavors, which does not lead to a local minimum. In 
Section \ref{extension} we describe an extension of SQCD with massless 
flavors, discuss its dynamics, and study the introduction of massive 
flavors, which in this case lead to a non-trivial, SUSY breaking minimum. We 
show that this minimum can be made parametrically long-lived.

Section \ref{review} provides mostly background material. Properties 
of fractional branes and their quiver gauge theories are sketched in
Section \ref{generalreview}. Section \ref{toiss} motivates focusing on
fractional D-branes at obstructed geometries. Section \ref{rundp1} 
reviews the dynamics of the simplest example in this class, the $dP_1$ 
theory, in 
the absence of extra massive flavors.

Section \ref{flavdp1} considers the quiver theory arising on fractional 
branes on the complex cone over $dP_1$ when massive fundamental flavors are 
added from a purely field theoretic perspective. This model 
is almost identical to the one in Section \ref{extension} and we show that 
it has a meta-stable SUSY breaking minimum. We also show that the minimum
can be parametrically long-lived. Before the addition of 
massive flavors, this theory is the simplest example of dynamical SUSY 
breaking (with runaway to infinite field values) due to obstructed 
deformation.

Section \ref{stringconstr} explains how to engineer the gauge theory of 
Section \ref{flavdp1} using D-branes in the complex cone over $dP_1$. 
The additional fundamental flavors are incorporated by introducing 
D7-branes at the singular geometry, while the flavor masses correspond to 
suitable vevs for D7-D7 scalars.

Appendix \ref{comput} describes the computation of the one-loop 
potential for classically flat directions in order to verify the 
existence or not of local SUSY breaking minima in the different theories 
we consider. Appendix \ref{section_meta_toy} considers SQCD with massless 
and additional massive flavors, Appendix \ref{section_meta_extended}
studies the extended version, and Appendix \ref{section_meta_dP1} 
describes the computation for the $dP_1$ theory.

Appendix \ref{D7branes} develops a general method to construct a 
class of D7-branes wrapping holomorphic 4-cycles in generic toric 
singularities and to identify their effect in the gauge theories on probe 
or fractional D3-branes. Appendix \ref{D7dp0} describes the construction 
for the complex cone over $dP_0$, where it can compared with orbifold 
techniques, since the geometry is $\IC^3/\IZ_3$. The rules are generalized 
in Appendix \ref{generatoric}, and applied for the complex cone over 
$dP_1$ in Appendix \ref{d7fordp1}.

\section{Meta-stable vacua in $\mathcal{N}=1$ SQCD-like theories with massive and 
massless flavors}

\label{sqcd}

In this section we first review the analysis in \cite{Intriligator:2006dd}
to determine the existence of meta-stable vacua in $\mathcal{N}=1$ $SU(N_c)$ SYM 
with massive flavors, and then generalize it to $SU(N_c)$ SQCD with 
massless and massive flavors. We show that this theory does not have
a meta-stable SUSY breaking minimum. We then construct a simple 
modification of the model that possesses a meta-stable SUSY breaking 
minimum. This model constitutes an interesting proposal for SUSY breaking 
in theories with massless flavors. In addition, it will be an extremely 
useful toy model of more involved quiver gauge theories arising from 
D3-branes at singularities in Section \ref{flavdp1}.

\subsection{$\mathcal{N}=1$ SQCD with light massive flavors}

\label{puresqcd}

Let us recall the system studied in \cite{Intriligator:2006dd}. Consider 
$SU(N_c)$ SYM with $N_f$ massive flavors $Q$, ${\tilde Q}$ with mass much 
smaller than $\Lambda_{SQCD}$, the dynamical scale of the gauge theory. 
We consider the flavor fields to have canonical K\"ahler potential.

The superpotential of the electric theory, for the case of equal flavor 
masses, is
\beqa
W\, = \, m\, \tr {\tilde Q} Q
\eeqa
In order to have an IR free dual description, so that the K\"ahler 
potential is under control in the small field region, we require  
\footnote{The possibility of extending the conclusions presented in this 
section outside of this range has been contemplated in 
\cite{Intriligator:2006dd}.}
$N_c+1 \leq N_f < {3\over 2} N_c$. 

The dual theory is $SU(N)$ SYM with $N=N_f-N_c$, with $N_f$ flavors $q$, 
${\tilde q}$, and mesons $M$. They transform as $(\fund,\antifund,1)$, 
$(\antifund, 1, \fund)$ and $(1,\fund,\antifund)$ under the 
$SU(N)\times SU(N_f)\times SU(N_f)$ color and flavor symmetry.
The superpotential for the dual theory is
\beqa
W\, =\, \frac{1}{\hat\Lambda} \, \Tr M q {\tilde q} \, +\, m\, \Tr M
\eeqa
where $\hat\Lambda$ is related to the dynamical scale $\Lambda_{SQCD}$ of 
the electric theory and $\Lambda$ of the magnetic theory by
\beqa
\Lambda_{SQCD}^{\, 3N_c-N_f}\, \Lambda^{3(N_f-N_c)-N_f}\, =\, 
\hat \Lambda^{N_f}
\label{match}
\eeqa
By redefining\footnote{In this simplified discussion, we ignore 
possible normalization factors in the K\"ahler potential of the fields. 
They can be nevertheless absorbed in additional redefinitions of 
flavor fields, mesons, and couplings, see  \cite{Intriligator:2006dd} 
for details. A similar comment applies to later examples.} the mesons as 
$\Phi= M/\Lambda$ and introducing the couplings $h=\Lambda/{\hat\Lambda}$, 
$\mu^2=-m\hat\Lambda$, the superpotential is of the form
\beqa
W\, =\, h \, \Tr\, q\, \Phi\, {\tilde q}\, -\, h\mu^2 \Tr \, \Phi
\label{W_SQCD_mag_1}
\eeqa
(where the traces run over flavor indices). 
Notice that for $N_f=N_c+1$ some further discussion is needed to establish 
that this superpotential correctly describes the effective dynamics, see 
\cite{Intriligator:2006dd} for details. A similar comment applies to all 
our forthcoming theories.

This theory breaks supersymmetry at tree level, since the F-flatness for 
$\Phi$ requires
\beq
{\tilde q}^i\, q_{j} \, = \, \mu^2 \, \delta^i_{\, j} 
\eeq
which cannot be satisfied, given that the rank of $\delta^i_{\, j}$ is 
$N_f$ while the rank \footnote{Since the theory is IR free, the rank of 
${\tilde q}^i\, q_{j}$ corresponds to its  classical value.} 
of ${\tilde q}^i\, q_{j}$ is $N<N_f$. This mechanism for spontaneous 
SUSY breaking at tree-level has been dubbed the {\bf rank condition 
mechanism} in \cite{Intriligator:2006dd}. There is a classical moduli space 
of minima with $V_{min}=(N_f-N)|h^2 \mu^4|$, parametrized by the vevs
\beqa
\Phi\, =\, \pmatrix{ 0 & 0\cr 0 & \Phi_0}
\qquad q=\pmatrix{\varphi_0\cr
0},\qquad \tilde q^T=\pmatrix{\tilde \varphi_0\cr 0},\qquad
{\rm with}\,\,\,\,
\tilde\varphi_0\varphi_0 = \mu^2\id_{N}.
\label{classpure}
\eeqa
A careful analytical computation shows that all pseudomoduli (classical 
flat directions not corresponding to Goldstone directions) are lifted by 
the one-loop effective potential, and that the maximally symmetric point 
in the classical moduli 
space
\beqa
\Phi_0=0,\qquad \varphi_0=\tilde \varphi_0=\mu\id_{N},
\eeqa
is a minimum of the one-loop effective potential. The one-loop effective 
potential at a generic point in the classical moduli space (\ref{classpure}) 
is the Coleman-Weinberg potential induced by the massive fluctuations 
around that point. We refer the reader to \cite{Intriligator:2006dd} 
for additional details, and to Appendix \ref{comput} for similar 
computations (in more involved situations).

The $SU(N)$ gauge dynamics is IR free and hence not relevant in the small 
field region, but it is crucial in the large field region. In fact, it 
leads to the appearance of the $N_f-N$ supersymmetric vacua predicted by 
the Witten index in the electric theory. In the region of large $\Phi$ 
vevs, $|\mu|\ll |\langle h\Phi \rangle|$, the $N_f$ 
flavors get large masses due to the cubic coupling in 
\eref{W_SQCD_mag_1}, and we recover pure $SU(N)$ SYM dynamics, with a 
dynamical scale $\Lambda'$ given by
\beqa
\Lambda'^{3N} \, =\, \frac{h^{N_f} \, \det \Phi}{\Lambda^{N_f-3N}} 
\eeqa
where $\Lambda$ is the Landau pole scale of the IR free theory. The 
complete superpotential, including the non-perturbative $SU(N)$ contribution
is
\beqa
W = N\, (\, h^{N_f}\, \Lambda^{-(N_f-3N)}\, \det\,\Phi\, )^{1/{N}}\, -\, h
 \mu^2\, \Tr\,\Phi
\eeqa
This superpotential leads to $N_f-N$ supersymmetric minima at
\beqa
& \langle h \Phi\rangle \, =\, \Lambda\, \epsilon^{\frac{2N}{N_f-N}} \,
\id_{N_f} =\, \mu \epsilon^{-\frac{N_f-3N}{N_f-N}}\,\id_{N_f}
\eeqa
where $\epsilon \equiv {\mu \over \Lambda}$. In the regime $\epsilon\ll 
1$, the vevs are much smaller than the Landau pole scale, and the analysis 
can be trusted. Notice also that these minima sit at $\left| \langle h\Phi 
\rangle \right| \gg \left| \mu \right|$, hence at a very large distance in 
field space from the local non-supersymmetric minimum. This large 
distance, in conjunction with the height of the potential barrier 
separating them from the non-SUSY minimum (which can be estimated from 
the classical superpotential) determines that the SUSY breaking 
meta-stable minimum is parametrically long-lived 
\cite{Intriligator:2006dd}.

\subsection{$\mathcal{N}=1$ SQCD with massless and massive flavors}

\label{flavoursqcd}

In this section we extend the previous discussion about meta-stable vacua in 
$\mathcal{N}=1$ SQCD with massive flavors \cite{Intriligator:2006dd} to another 
system. We investigate the case in which, in addition to light massive flavors,
there are massless flavors. We consider $SU(N_c)$ SQCD with $N_{f,0}$ 
massless flavors ${\tilde Q}_0$, $Q_{0}$ and $N_{f,1}$ massive flavors 
${\tilde Q}_1$, $Q_1$, with mass much smaller than $\Lambda_{SQCD}$.
Again, we consider canonical K\"ahler potential for these fields.
To simplify the expressions, flavor indices are kept implicit.
The superpotential, for the equal mass case, is thus
\beqa
W\, =\, m\, \Tr {\tilde Q}_1 Q_1
\eeqa

As before, in order to have control over the computations in the IR, we 
consider the theory in the free magnetic range
$N_c+1 \leq N_f=N_{f,0}+N_{f,1}< {3\over 2} N_c$.
In order for the classical theory to have SUSY breaking due to rank 
condition mechanism at tree level, we further require
\beq
N_{f,1} > N = N_{f,0}+N_{f,1}-N_c \ \ \ \ \Leftrightarrow \ \ \ \ N_c 
> N_{f,0}
\label{obstr}
\eeq
This condition is interesting, and will reappear in Sections 
\ref{toiss} and \ref{flavdp1} in the context of branes at singularities.
We now study this theory in detail since it is natural to ask whether a 
SUSY breaking meta-stable minimum exists. It will also serve as a 
warm-up for the modified model of Section \ref{extension}.

The dual magnetic theory is $SU(N)$ SQCD with $N=N_f-N_c$, and dynamical 
scale $\Lambda$. There are $N_f=N_{f,0}+N_{f,1}$ flavors ${\tilde q}_0$, 
$q_{0}$ and ${\tilde q}_1$, $q_{1}$, and the mesons $\Phi_{00}$, 
$\Phi_{01}$, $\Phi_{10}$, $\Phi_{11}$. The latter can be expressed as 
composites of the electric theory, namely $\Phi_{ij}=\frac 1\Lambda 
\tilde{Q}_i Q_j$. The complete superpotential, in 
the limit where the $SU(N)$ dynamics is ignored, can be written
\beqa
W & = & h\, \Tr q\, \Phi\, {\tilde q}\, - \, h \mu^2\, \Tr \Phi_{11}
\label{supoflavoursqcd}
\eeqa
where $h$ and $\mu$ are defined as in the theory without massless flavors 
in Section \ref{puresqcd}. Notice that for simplicity we have taken the 
coupling constants of the cubic terms involving $\Phi_{ij}$, $i,j=0,1$, to 
be equal, even though no global symmetry imposes that restriction.

The equations of motion for $\Phi_{11}$ are
\beqa
{\tilde q}_1^i\, q_{1,j} \, = \, \mu^2 \, \delta^i_{\, j} 
\eeqa
Since $N_{f,1}-N=N_{f,1}-(N_{f,1}+N_{f,0}-N_c)=N_c-N_{f,0}>0$, there is 
SUSY breaking by the rank condition.

\subsubsection*{Absence of a local minimum}
\label{nominimum}

A detailed analysis of moduli and pseudomoduli and the computation of 
their masses is provided in Appendix \ref{section_meta_toy}. We summarize 
the results here.

There is a classical moduli space of degenerate supersymmetry breaking minima
with $V_{min}=(N_{f,1}-N)|h^2 \mu^4|$. 
This classical moduli space can be parametrized as follows
\beq
\begin{array}{ccccc}
q_0={\tilde q}_0=0 & \ \ \ \ & q_1 \, =\, \pmatrix{\, \varphi_1\, ;\, 0\,} & \ \ \ \ &
{\tilde q}_1 \, =\, \pmatrix{\, {\tilde \varphi}_1\, \cr \, 0\,} \\ \\
\Phi_{01}\, =\, \pmatrix{ \, 0 \, ;\, Y\, } & &
\Phi_{10}\, = \, \pmatrix{\, 0\, \cr \, {\tilde Y} \,}  & &
\Phi_{11}\, =\, \pmatrix{\, 0\,\,\, 0 \, \cr \, 0 \, X_1 \, } 
\end{array}
\label{parametrization_toy_model}
\eeq
where ${\tilde \varphi}_1$, $\varphi_1$ are $N\times N$ blocks satisfying
${\tilde \varphi}_1\varphi_1=\mu^2 \, \id_N$. In addition, 
$Y$, ${\tilde Y}$ and $X_1$ are $N_{f,0}\times (N_{f,1}-N)$,
$(N_{f,1}-N)\times N_{f,0}$ and $(N_{f,1}-N)\times (N_{f,1}-N)$ blocks, 
respectively. The vev for $\Phi_{00}=X_0$ is arbitrary. 

Goldstone bosons corresponding to broken global symmetries remain exactly 
massless. Integrating out classically massive fluctuations, the one-loop 
effective potential becomes

\beq
\langle V_{eff}^{(1)}\rangle=const.+|h^4 \mu^2| \ {(\log 4-1)\over 16 \pi^2}N\left[(N_f-N)\left(2 \ |\delta \Phi_1|^2+|\mu^2| (\theta+\theta^*)^2\right)+\tilde{N}\left(|\delta Y|^2+|\delta \tilde{Y}|^2 \right)\right]+\ldots
\eeq

The variables $\theta$, etc are defined in \eref{expansion_toy_model} and 
$\tilde{N}=\min(N_{f,0},N_{f,1}-N)$.
This expression assumes that the non-vanishing parts of $Y$ and $\tilde{Y}$ 
are proportional to the identity. See \eref{m1} for a slightly more 
general equation.

We see that $\delta \Phi_0$ remains massless at 1--loop. In principle, it 
is still possible that $\delta \Phi_0$ becomes massive at higher loops, 
producing a meta-stable minimum (probably with a much smaller potential 
barrier) at small expectation values for the fields. We will not consider 
this possibility, but will explore an extension of this model for which 
this flat direction is lifted at the classical level in next section.

\subsubsection*{Behavior at large fields}

In analogy with the case in Section \ref{puresqcd}, we expect that
in the large field region we recover the low-energy structure of the 
$SU(N_c)$ theory with $N_{f,0}<N_c$ flavors, namely an Affleck-Dine-Seiberg 
(ADS) superpotential triggering a runaway behavior for the meson 
$\Phi_0={\tilde Q}_0 Q_0$ of the electric theory. 
This is indeed the case, and we now show it from the perspective of the magnetic theory.

Consider a generic point in the moduli space of the magnetic theory with non-vanishing 
expectation values of $\Phi_{00}$ and $\Phi_{11}$. The flavors 
${\tilde q}_0$ 
and $q_0$ become massive, with mass matrix $h \Phi_{00}$. Similarly, 
${\tilde q}_1$ and $q_1$ get masses given by $h \Phi_{11}$. We can then 
integrate out these fields, solving their equations of motion by setting 
$\Phi_{10}=\Phi_{01}=0$. The resulting theory is pure $SU(N)$ with a 
dynamical scale given by

\beqa
\Lambda'^{\, 3N}\, = \, \frac{h^{N_f} \det\Phi_{00} 
\det\Phi_1}{\Lambda^{N_f-3N}}
\eeqa

The complete superpotential, including the non-perturbative piece, reads
\beqa
W & = & -\, h\mu^2 \Tr \Phi_{11} \, +\, 
N\, \left(\, \frac{h^{N_f} \, \det\Phi_{00} \, 
\det\Phi_{11}}{\Lambda^{N_f-3N}}\, \right)^{\frac{1}{N}}
\label{supoall}
\eeqa

Recall that $\Phi_{00}$ is (up to a rescaling by $1/\Lambda$) equal to the meson 
$\Phi_0$ of the original electric theory, so 
we are interested in its dynamics. The effective action for 
$\Phi_{00}$ can be obtained by using the equation of motion for 
$\Phi_{11}$. Integrating out $\Phi_{11}$ we obtain

\beqa
W & = & -(N_{f,1}-N)\,\, \left(\frac{\mu^{2N_{f,1}}\, 
\Lambda^{N_f-3N}}{h^{N_{f,0}}\, \det 
\Phi_{00}}\right)^{\frac{1}{N_{f,1}-N}}
\label{final}
\eeqa
Noticing that $N_{f,1}-N=N_c-N_{f,0}$ in terms of parameters of the 
electric theory, this is exactly the runaway superpotential for $\Phi_0$ 
induced by $SU(N_c)$ SQCD with $N_{f,0}$ massless flavors. The unfamiliar 
structure of factors inside the bracket is simply due to the fact that 
the $SU(N_c)$ dynamical scale appears expressed in terms of the Landau 
pole scale of the magnetic theory. 

Notice that the runaway potential can be trusted as long as $|h\langle 
\Phi_{00} \rangle|\ll \Lambda$. For larger fields, the electric theory 
completes the UV and ensures that the runaway persists to infinity.

\medskip

As described in Section \ref{nominimum} and Appendix 
\ref{section_meta_toy}, the 
one-loop potential for this theory leaves the field $\Phi_{00}$ massless 
around the origin, and moreover slopes down as this field increases.
Hence, $\Phi_{00}$ is not stabilized in the small vev region. 
At large fields, we have a runaway to infinity for this field. The most 
conservative proposal is thus to connect these two behaviors in a 
constantly 
decreasing potential in the direction $\Phi_{00}$. Of course higher loop 
contributions could in principle lead to a non-trivial behavior in the 
intermediate field regime. Since this is however difficult to establish, 
in the next section we turn to the study of a different theory, which 
incorporates a mild extension of the above model. We will show that 
this new theory does have a meta-stable minimum  separated from a runaway 
behavior at infinity by a potential barrier. In addition, 
the extension brings the model closer to quiver gauge theories with 
obstructed deformations, which are studied in later sections.

\subsection{Extension of SQCD with massless flavors}
\label{extension}

As discussed in the previous section, SQCD with massless and massive flavors 
does not have a meta-stable SUSY breaking minimum at one-loop. In this section we 
propose a simple extension of this theory that does have such 
meta-stable SUSY breaking minimum, which is separated  from a runaway 
behavior at infinity by a potential barrier and is parametrically 
long-lived. 

The extension amounts to the introduction of a new field $\Sigma_0$, with 
cubic coupling to the flavors of the original electric theory. The role 
of this extension in leading to a meta-stable vacuum is easily understood. 
In the dual theory, the field $\Sigma_0$ couples to the meson $\Phi_{00}$ 
via a mass term, forcing the vev of the latter to vanish. Hence the new 
term eliminates the $\Phi_{00}$ direction which was not properly lifted by 
the one-loop potential.

\subsubsection*{The extended theory without massive flavors}
\label{extwithout}

Let us start by describing the extended theory and its dynamics in the 
absence of massive flavors. Consider $SU(N_c)$ SQCD with $N_{f,0}<N_c$ 
massless flavors ${\tilde Q}_0$, $Q_{0}$ and add a set of singlets 
$\Sigma_0$, transforming in the bi-fundamental of the $SU(N_{f,0})^2$ 
flavor global symmetry, with a superpotential
\beqa
W_{ext}\, =\, g\, \Tr Q_0\, \Sigma_0\, {\tilde Q}_0
\label{extsupo}
\eeqa
where $g$ is a dimensionless coupling \footnote{This model has been 
discussed in \cite{Barnes:2004jj}, where it was called SSQCD (for singlets + SQCD).
In that paper, the IR phases of this theory were studied using Seiberg duality and 
a-maximization.}. We consider canonical K\"ahler 
potentials for all fields.

This theory has a runaway behavior in the new field $\Sigma_0$. 
To show this, we introduce the gauge invariant mesons\footnote{Although 
almost identical, this meson differs slightly from the $\Phi_{00}$ we 
defined in Section \ref{flavoursqcd}. The latter was an elementary field 
in the magnetic theory, so we included a power of $\Lambda$ in its
definition to give it canonical dimensions.}  
$\Phi_{0}=Q_0{\tilde Q_0}$. The complete superpotential is
\beqa
W\, =\, g\, \Tr \Sigma_0 \Phi_{0} + (N_c-N_{f,0})\, \left(
\, \frac{\Lambda_{SQCD}^{3N_c-N_{f,0}}}{\det \Phi_{0}} \, 
\right)^{\frac{1}{N_{f,0}-N_c}}
\eeqa
Upon using the equation of motion for $\Phi_{0}$ we have
\beqa
W\, =\, N_c\, \left(\, g^{\,N_{f,0}}\, \Lambda_{SQCD}^{3N_c-N_{f,0}}\, 
\det\Sigma_0\, \right)^{\frac{1}{N_c}}
\label{runsqcd}
\eeqa
This is indeed a runaway behavior: the F-term for $\Sigma_0$ gives
\beqa
\frac{\partial W}{\partial (\Sigma_0)_{ij}}\, \simeq \, 
(\det\Sigma_0)^{1/N_c}\, (\Sigma_0^{-1})_{ji}
\eeqa
which, scaling $\Sigma_0\to \lambda \Sigma_0$, scales as
$\lambda^{\frac{N_{f,0}-N_c}{N_c}}$. Hence since $N_{f,0}<N_c$, the 
F-terms relax to zero for large fields. With some foresight, we note that 
this behavior is completely analogous to the one we will discuss in 
Section \ref{rundp1} for the $dP_1$ quiver gauge theory.

\subsubsection*{Introducing massive flavors in the extended theory}

The above theory is a close cousin of SQCD, and in particular it 
shares its runaway behavior (albeit in a different field direction). 
However, as we discuss later, they differ in their dynamics when extra 
light massive flavors are introduced. In particular, the extended 
theory will show meta-stable minima.

The extended theory with massive flavors is a combination of the SQCD 
with massless and massive flavors of Section \ref{flavoursqcd} and the 
extension term introduced above. Hence we consider $SU(N_c)$ SQCD 
with $N_{f,0}$, $N_{f,1}$ massless and massive flavors, and fields 
$\Sigma_0$, coupled to the massless flavors via (\ref{extsupo}).
As in Section \ref{flavoursqcd}, we consider $N_{f,0}<N_c$ and hence 
$N_{f,1}>N$, so that the dual theory has supersymmetry breaking by the 
rank condition at tree level.

The dual magnetic theory is $SU(N)$ SQCD with $N=N_f-N_c$, and
$N_f=N_{f,0}+N_{f,1}$ flavors, with dynamical scale $\Lambda$. We also 
have mesons $\Phi_{ij}=\frac{1}{\Lambda} {\tilde Q}_iQ_j$, and 
the classical superpotential
\beqa
W & = & h\, \Tr q\, \Phi\, {\tilde q}\, - \, h \mu^2\, \Tr 
\Phi_{11}\, +\, h \mu_0\Tr \Sigma_0 \Phi_{00}
\label{supoextsqcd}
\eeqa
where $h=\Lambda/\hat\Lambda$, $\mu^2=-m\hat\Lambda$, $\mu_0=g\Lambda$, 
and $\hat\Lambda$, $\Lambda$ are related to the electric scale 
$\Lambda_{SQCD}$ by (\ref{match}).

As usual the equations of motion for $\Phi_{11}$ lead to SUSY breaking by 
the rank condition.

\subsubsection*{The local minimum}

Although \eref{extsupo} is a simple modification of (\ref{supoflavoursqcd}), 
the addition of the new field and its interactions has a drastic effect in 
the small field region of the theory. A full discussion of pseudomoduli and
their masses in this theory is given in Appendix \ref{section_meta_extended}. 
The classical SUSY breaking minima are parametrized as in \eref{parametrization_toy_model}, with 
$X_0$ fixed to zero by the tree-level superpotential (i.e. $X_0$ is no
longer a pseudomodulus).

The one-loop effective potential has a critical point at 
$\Phi_1=Y=\tilde{Y}=(\theta+\theta^*)=0$. Around this point, it becomes  

\beq
\langle V_{eff}^{(1)}\rangle=const.+|h^4 \mu^2| \ {(\log 4-1)\over 16 \pi^2}N\left[(N_f-N)\left(2 \ |\delta \Phi_1|^2+|\mu^2| (\theta+\theta^*)^2\right)+\tilde{N}\left(|\delta Y|^2+|\delta \tilde{Y}|^2 \right)\right]+\ldots
\eeq

Hence, all pseudomoduli get positive masses and are thus stabilized. The 
critical point becomes a meta-stable minimum, whose longevity we analyze later.
Again the expression above corresponds to the non-vanishing parts of $Y$ 
and $\tilde{Y}$ being proportional to the identity. Equation \eref{m2} 
gives the general result.

\subsubsection*{Behavior at large fields}

In the large field region of the theory \eref{supoextsqcd}, we expect to 
recover the behavior in the absence of massive flavors. Namely, we expect 
a runaway of $\Sigma_0$ dictated by (\ref{runsqcd}). This is indeed 
the case as we now show.

We can study the large field region by considering the expression 
(\ref{final}), which describes the large field behavior for the 
non-extended theory, and adding the extension term $-h \mu_0 \Tr \Phi_0 
\Sigma_0$. Upon integrating out $\Phi_{00}$, we obtain

\beqa
W\, =\, -(N_f-N) \,\, 
\left( \, 
\mu_0^{N_{f,0}} \mu^{2N_{f,1}}\, \Lambda^{N_f-3N}\, 
\det \Sigma_0 \,\right)^{\frac{1}{N_f-N}}
\label{W_runaway_extended_model}
\eeqa

Recalling that $N_f-N=N_c$ in terms of the underlying electric theory, 
this behavior 
\footnote{Notice that, despite the fact that $\Phi_{00},\Phi_{11}\to 0$ as 
$\Sigma_0$ runs to infinity, for each fixed value of $\Sigma_0$ the flavor 
masses remain parametrically larger (for $\epsilon=\mu/\Lambda\ll 1$, 
$\alpha_\mu=\mu_0/\mu\gg 1$) than the vacuum energy, hence it is 
consistent to keep them integrated out, as implicitly done in our 
computation.} 
is essentially identical to the runaway of the extended 
theory without the massive flavors (\ref{runsqcd}). The different factors 
inside the bracket are simply due to expressing the superpotential in 
terms of the Landau pole scale of the magnetic theory.

\subsubsection*{Lifetime of meta-stable vacua}

The decay rate is proportional to the semi-classical decay probability. 
This probability is proportional to $\exp(-S)$, where the bounce action 
$S$ is the difference in the Euclidean action between
the tunneling configuration and the meta-stable vacuum.

The SUSY breaking, meta-stable vacuum is given by
\beq
q_1 \, =\, \pmatrix{\, \varphi_1\, ;\, 0\,}  \ \ \ \ 
{\tilde q}_1 \, =\, \pmatrix{\, {\tilde \varphi}_1\, \cr \, 0\,}
\eeq
with ${\tilde \varphi}_1=\varphi_1=\mu \, \id_N$, and all the other fields
having a zero expectation value.

We saw in the previous section that the SUSY vacuum corresponds to $\Sigma_0$ 
running away to infinity due to the superpotential 
\eref{W_runaway_extended_model}. Simultaneously, the equations of motion 
force the fields $\Phi_{00}$ and $\Phi_{11}$ to have non-zero vevs, 
adjusted to the $\Sigma_0$ vev.

In order to estimate the bounce action, we must find a trajectory in 
field space connecting the meta-stable and SUSY vacua such that the 
potential barrier is minimum. The classical superpotential \eref{supoextsqcd} does not have any coupling that would give rise to contributions
to the classical scalar potential of the form $|h^2 \varphi_1 \Sigma_0|^2$ or $|h^2 {\tilde \varphi}_1 \Sigma_0|^2$. 
Such contributions would become very large as $\Sigma_0$ runs away
if $\varphi_1$ or ${\tilde \varphi}_1$ do not vanish. Anyway, it is convenient 
to consider the following simple trajectory connecting both vacua, which exhibits the characteristic
barrier height and distance in field space separating the meta-stable and SUSY vacua: first go to the origin, where the
potential is

\beq
V_T=N_{f,1} \, |h^2 \mu^4|
\eeq
and then approach the SUSY vacuum at infinity increasing $\Sigma_0$. 

\begin{figure}[ht]
  \epsfxsize = 6cm
  \centerline{\epsfbox{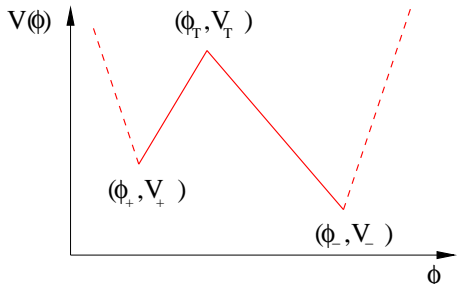}}
  \caption{A triangular potential barrier.}
  \label{triangular_barrier}
\end{figure}

To estimate $S$, we model the potential as a triangular barrier, for which the exact bounce
action has been derived in \cite{Duncan:1992ai}. A triangular barrier has the general form depicted in 
\fref{triangular_barrier}.

We define the quantities $\Delta \phi_\pm=\pm (\phi_T-\phi_\pm)$ and $\Delta V_{\pm}=(V_T-V_\pm)$.
A triangular barrier is a good approximation in cases in which the gradient of the potential
is approximately constant at both sides of the peak. In this case,

\beq
S \sim {\left|(\Delta \phi_+)^2 - (\Delta \phi_-)^2\right|^2\over \Delta V_+}
\label{S_triangular_barrier}
\eeq

\begin{figure}[ht]
  \epsfxsize = 7cm
  \centerline{\epsfbox{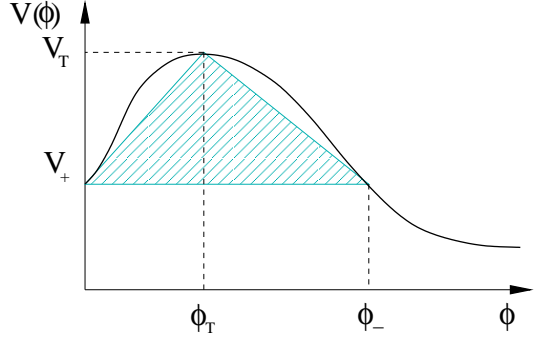}}
  \caption{Sketch of the potential along the bounce trajectory and the triangular barrier we use to model it.}
  \label{potential_barrier}
\end{figure}

Modeling the barrier should be done in slightly different way from the SQCD with 
light flavors case \cite{Intriligator:2006dd}, since in this case there is a runaway and the 
potential does not vanish at finite values of the fields. The slope of our potential becomes progressively
smaller as $\Sigma_0 \rightarrow \infty$. A cartoon of the potential is presented
in \fref{potential_barrier}, showing the criterion we will use to define the triangular barrier.

We should interpret the variable $\phi$ in \fref{potential_barrier} as parametrizing the trajectory in 
field space connecting the meta-stable and SUSY vacua. Hence, the region before the peak
corresponds to motion in $\varphi_1$ and $\tilde{\varphi}_1$, and the region after it corresponds
to motion in the ${\Sigma_0}$ direction. In addition,  \fref{potential_barrier} is not drawn at scale, and the distance
between the meta-stable minimum and the peak is negligible with respect 
to the rest of the plot.

Since the potential does not vanish, but asymptotes zero, we define $\phi_-$ 
as the point in the large field region at which the potential 
falls below $|h^2 \mu^4|$. The height of the barrier is also of order $|h^2 \mu^4|$. At the level of the
estimations we are making, the calculation we perform would also correspond to an alternative 
criterion: assuming that $V(\phi_-)=0$ (for $\phi_-$ as just defined) thus underestimating the
barrier and producing a lower bound for the bounce action.

Taking the simple ansatz $\Sigma_0=\sigma_0 \id_{N_{f,0}}$, we obtain

\beq
V \sim \left|{\mu^{2 N_{f,1}}\mu_0^{N_{f,0}} \Lambda^{N_f -3N} 
\sigma_0^{N-N_{f,1}}}\right|^{2\over N_f -N}
\eeq

Then, in order to have $V \sim |h^2 \mu^4|$, we have  
\beq
|\sigma_{0-}| \sim \left|\Lambda^{N_f-3N} \mu_0^{N_{f,0}} \over h^{N_f-N} \mu^{2(N_{f,0}-N)}\right|^{1\over N_{f,1}-N}
\eeq

In our case, $\Delta \phi_- \sim \mu$ is negligible, $\Delta V_+ \sim |h^2 \mu^4|$ and $\Delta \phi_- \sim \sigma_{0-}$. Using \eref{S_triangular_barrier}
we obtain

\beq
S \sim {1\over |h|^{6+4 N_{f,0}/(N_f-N)}} \ {|\alpha_\mu|  ^{4 N_{f,0}/ (N_{f,1}-N)}\over |\epsilon|^{4(N_f-3N)/ (N_{f,1}-N)}}
\label{bounce_S_extended_model}
\eeq
where we have defined $\epsilon=\mu/\Lambda$ as in Section \ref{puresqcd} and $\alpha_\mu=\mu_0/\mu$ measures the strength of the extension term
relative to the rest of the superpotential. The 
general behavior is clear: for a fixed value of the coupling $h$ the lifetime can be made parametrically large by either 
making the fundamental flavors in the original theory light (i.e. small $\epsilon$) or by increasing the relative strength 
of the extension term in the superpotential (given by $\alpha_\mu$).
A heuristic reason for the latter is that $\alpha_\mu$ indicates how much the extension term ``pushes'' the system into a runaway in 
the $\Sigma_0$ direction. 

For $N_{f,0}=0$ \eref{bounce_S_extended_model} becomes

\beq
S \sim {1 \over |h|^6} {1 \over |\epsilon|^{4(N_{f,1}-3N) / (N_{f,1}-N)}}
\eeq

The result of \cite{Intriligator:2006dd} is identical to this one except 
that the power of $|\epsilon|$ is $4(N_{f,1}-2N) / (N_{f,1}-N)$, i.e. larger,
in that case. This discrepancy is precisely accounted for by noticing 
that our criterion for determining the potential barrier 
underestimates $\Delta \phi_-$, and hence the bounce action, with respect 
to \cite{Intriligator:2006dd}.

\section{Review of fractional branes and obstructed deformations}

\label{review}

In this section we describe quiver gauge theories based on D3-branes at 
singularities 
with fractional branes. We point out that the gauge theories on the 
so-called `DSB fractional branes' have features analogous to the extended 
version of SQCD with massless flavors studied above. This will motivate 
the study of these theories with additional massive flavors in coming 
sections.

\subsection{General review of fractional branes}

\label{generalreview}

D-branes at singularities provide a useful arena to study and test the 
gauge/string correspondence, in situations with reduced (super)symmetry.
In particular, the introduction of fractional branes leads to interesting 
dual pairs involving non-conformal gauge theories with non-trivial 
dynamics in the infrared. In the string construction, fractional branes
correspond to D-branes wrapped on cycles collapsed at the singularity, 
consistently with cancellation of (local) RR tadpoles. At the level
of the gauge theory, fractional branes correspond to rank assignments for 
gauge factors in a way consistent with cancellation of non-abelian 
anomalies.

A particularly well-known class of systems corresponds to D3-branes at 
toric singularities. The corresponding gauge theories are described in 
terms of brane tiling or dimer graphs 
\cite{Hanany:2005ve,Franco:2005rj,Hanany:2005ss,Feng:2005gw,Franco:2006gc}.
We restrict to this class in what follows, although some facts are valid 
in non-toric singularities as well.

A classification of different kinds of fractional branes, the infrared 
behavior of the associated gauge theory, and corresponding features in 
the geometry, is as follows \cite{Franco:2005zu}: 

\bigskip

\noindent $\bullet$ {\bf $\NN=2$ fractional branes:} these are fractional 
branes 
whose quiver gauge theory (in the absence of D3-branes) corresponds to a 
closed loop of arrows passing through a set of nodes, with the associated 
gauge invariant operator {\em not} appearing in the superpotential. These 
fractional branes therefore have flat directions, parametrized by 
vevs for this mesonic operator, and along which the effective theory is 
$\NN=2$ supersymmetric. Geometrically, 
these fractional branes exist for singularities which are not isolated, 
but have (complex) curves of $\IC^2/\IZ_N$ singularities passing through 
them. The fractional branes correspond to D5-branes wrapped on the 
2-spheres collapsed at the latter. The prototypical example is provided by 
branes at the $\IC^2/\IZ_2$ singularity. In the gauge/gravity 
description, the IR dynamics of the gauge theory (instantons and 
Seiberg-Witten points) corresponds to an enhan\c{c}on behavior on the 
gravity side.

\medskip

\noindent $\bullet$ {\bf Deformation fractional branes:} these are 
fractional branes 
whose quiver is either given by a set of decoupled nodes, or by a set of 
nodes joined by a closed loop of arrows, with the corresponding gauge 
invariant operator appearing in the superpotential. Moreover, the 
involved gauge factors all have the same rank. 
Geometrically, these fractional branes are associated with a possible 
complex deformation of the singularity. These are easily described in 
terms of splitting of the web diagram \cite{Aharony:1997ju,Aharony:1997bh,Leung:1997tw} 
of the singularity into sub-webs. The prototypical 
example is provided by branes at the conifold singularity. The behavior 
of the gauge theory corresponds to confinement of the involved gauge 
groups, and in the dual gravity background this corresponds to a complex 
deformation leading to finite size 3-cycles.

\medskip

\noindent $\bullet$ {\bf DSB fractional branes:} these are fractional 
branes of 
any other kind, hence they provide the generic case. They are fractional 
branes for which the non-trivial gauge factors have 
unequal ranks. Geometrically, they are associated with obstructed 
geometries, which do not admit the corresponding complex deformation
\footnote{An important and often unnoticed fact, which has been discussed 
in \cite{Franco:2005zu}, is that geometries 
admitting complex deformations may have DSB fractional branes, since 
generically the number of complex deformations is smaller than the number 
of independent fractional branes. An example is provided by the complex 
cone over $dP_3$, which admits two complex deformations and three 
independent fractional branes.}. As discussed in 
\cite{Berenstein:2005xa,Franco:2005zu,Bertolini:2005di}, the dynamics 
of the gauge theory corresponds to the appearance of an ADS superpotential 
which removes the supersymmetric minimum. Moreover, as first discussed in 
\cite{Franco:2005zu} and later studied in detail in 
\cite{Intriligator:2005aw,Brini:2006ej}) the theory has a runaway behavior 
towards infinity (in a direction parametrized by di-baryonic operators), 
at least in the large field regime. The absence of a vacuum at 
finite values of the fields suggests that the dual supergravity 
background, describing the UV behavior of the theory may not admit a 
smoothing of their naked singularities. The prototypical case is the 
fractional brane of the complex cone over $dP_1$, which we 
study in next section.

\subsection{Quiver gauge theories and the ISS proposal}
\label{toiss}

We would like to consider the possible generalization of the ISS proposal 
to quiver gauge theories on fractional branes. Notice that adding massive 
flavors to $N=2$ SYM was shown in \cite{Intriligator:2006dd} not to lead 
to SUSY breaking local meta-stable minima. Hence $N=2$ fractional branes 
are not appropriate candidates to implement the ISS proposal. 

On the other hand, one can consider deformation branes leading to a 
set of decoupled $\mathcal{N}=1$ SYM theories in the infrared, the simplest case 
being the fractional brane of the conifold theory. Addition 
of flavors to these theories leads to a direct realization of the ISS 
model, so the analysis in \cite{Intriligator:2006dd} goes through without 
modification. Thus, these are the simplest examples of D-brane 
configurations realizing the ISS proposal in the D-brane. Notice however 
that deformation branes whose quiver gauge theory reduces simply to $\mathcal{N}=1$ 
SYM exist only for very non-generic cases, like vector-like theories.

Hence to understand the generic extension of the ISS proposal to quiver 
gauge theories we have to consider the remaining cases. They correspond to 
more general deformation fractional branes (leading to a set of nodes 
joined by arrows, like for instance the three-node fractional brane of 
the $dP_3$ theory studied in \cite{Franco:2005fd}), or DSB fractional 
branes. All these gauge theories contain massless 
bi-fundamentals, hence the relevant version of the ISS proposal is that 
provided in Section \ref{extension}. For the ISS proposal to have a 
chance to work, some necessary conditions are required. First,
the theory after dualizing the node in the free magnetic phase, should 
have SUSY breaking by the rank condition. As in Sections \ref{flavoursqcd}, 
\ref{extension}, this requires $N_{f,1}>N_c$, equivalently $N_{f,0}<N_c$. 
This condition is not satisfied by deformation fractional branes of the 
kind we are considering 
(namely, with quivers given by equal rank nodes joined by arrows). But it
is satisfied for DSB branes, on which we center henceforth. A second 
condition is the 
existence of suitable fields with cubic couplings to the massless flavors 
of the node in the free magnetic phase. It is easy to verify that in all 
known examples of DSB branes this condition is satisfied as well.

Hence DSB fractional branes are the natural setup to provide a 
generalization of the ISS proposal for generic quiver gauge theories.
In the remainder of the paper, we focus on systems of DSB fractional 
branes, in the absence of D3-branes. These theories can be regarded and 
studied on their own. Alternatively, one can consider them as the IR 
result of a duality cascade, along which the D3-branes have disappeared. 
Some remarks about the latter interpretation, and the related issue of 
gravity duals of the gauge theory phenomena we discuss, are presented in 
Section \ref{section_duality_cascade}.

\subsection{Review of the runaway for fractional branes in the cone over 
$dP_1$}

\label{rundp1}

We now focus on DSB fractional branes. In this section we review the 
dynamics in the absence of additional massive flavors, while in the next 
one we consider the possible appearance of meta-stable vacua once 
extra flavors are included. Both in this and coming sections we 
center on the simplest situation of the fractional brane of the complex 
cone over $dP_1$. However, we expect that much of our discussion 
is valid for the general case, and in fact our analysis is 
automatically valid for other 
examples of fractional branes on obstructed geometries leading to the 
same quiver gauge theory (for instance, the fractional branes of $dP_5$ 
studied in \cite{Diaconescu:2005pc}).

\begin{figure}[ht]
  \epsfxsize = 4.2cm
  \centerline{\epsfbox{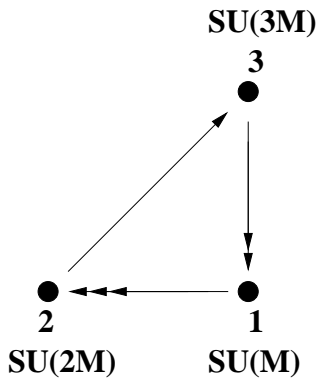}}
  \caption{Quiver diagram for $M$ fractional branes in the complex 
  cone over $dP_1$.}
  \label{dP1_quiver_D5}
\end{figure}

The gauge theory on D3-branes at a singularity given by a complex cone  
over $dP_1$ was determined in \cite{Feng:2000mi}. The gauge theory on a 
set of fractional branes has quiver shown in Figure \ref{dP1_quiver_D5}. 
For convenience we have 
labeled the nodes such that the gauge factor associated with a node with 
label $k$ is $SU(kM)$. Please note that this convention is different from 
others in the literature.

The superpotential is given by
\beqa
W & = &  \lambda\, (\, X_{23}X_{31}Y_{12}\, -\, X_{23} Y_{31} X_{12}\, )
\label{supod3}
\eeqa
with obvious notation. Here and in what follows, traces over color indices are implicit.
We have introduced a dimensionless coupling $\lambda$, which is equal for 
both terms since they are related by an $SU(2)$ global symmetry.

This theory develops a non-perturbative superpotential which removes the 
supersymmetric minimum 
\cite{Berenstein:2005xa,Franco:2005zu,Bertolini:2005di}. Indeed, the 
theory has a runaway towards infinity \footnote{Notice that this statement 
requires some assumption about the K\"ahler potential for the $dP_1$ theory.} 
\cite{Franco:2005zu}, see 
\cite{Intriligator:2005aw} for a detailed discussion. 
The runaway can be easily seen in both the $SU(2M)$ and $SU(3M)$
dominated regimes. The computation is very similar in both cases, and we now review 
the situation in which the $SU(3M)$ dynamics 
dominates. Since this gauge factor confines, we construct 
the mesons
\beqa
M_{21}\, = \, X_{23}X_{31} \quad ; \quad M_{21}'\, =\, X_{23} Y_{31} 
\eeqa
The $SU(3M)$ gauge factor has $2M$ flavors, and leads to a 
non-perturbative ADS superpotential. The full superpotential is 
\beqa
W & = &  \lambda \, (\,M_{21}Y_{12}\, -\, M_{21}' X_{12}\,)\, + \,
M \, \left( \, \frac{\Lambda_3^{7M}}{\det{\cal M}}\,\right)^{\frac 1M}
\label{unfldp1}
\eeqa
where ${\cal M}=(M_{21};M_{21}')$ is the mesonic $2M\times 2M$ matrix.

For simplicity let us focus on the case $M=1$, and denote
\beqa
M_{21}\, =\, \pmatrix{A \cr C} \quad ; \quad M_{21}' \, =\, \pmatrix{B 
\cr D} \quad ; \quad Y_{12}\, =\, (a, b) \quad ; \quad X_{12}\, =\, (c,d)
\eeqa
We then have
\beqa
W & = & \lambda\, (\, aA\,+\,bC\, -\, cB\, -\, dD \,)\, + \,
\frac{\Lambda_3^{7}}{AD\, - \, BC}
\eeqa
Using the equations of motion for $A$, $B$, $C$, $D$ we obtain
\beqa
W & \simeq & (\, \lambda^2 \,\Lambda_3^{\, 7} \,  \det{\cal Y}\, )^{1/3}
\eeqa
where ${\cal Y}=\pmatrix{a & b \cr c & d}=\pmatrix{Y_{12}\cr X_{12}}$ and we have dropped
an unimportant numerical factor.
This superpotential leads to a runaway behavior \footnote{Again, notice that the 
existence of a runaway scalar potential implies certain assumptions for 
the K\"ahler potential.} for the fields $X_{12}$, 
$Y_{12}$. Notice that these fields are not the mesons of the confining group, but 
rather the microscopic fields which had cubic couplings with the original 
flavors. 

Along this direction in field space, the additional gauge symmetry 
$SU(2M)\times SU(M)$ is generically Higgssed by the vevs for ${\cal Y}$, 
hence it does not lead to any modifications of the above behavior.

The conclusion is that the theory has a runaway in the direction 
${\cal Y}$ corresponding to the singlets (of the strong dynamics gauge 
factor) with cubic couplings with the flavors $X_{32}$, $X_{13}$, 
$Y_{13}$. This behavior is reminiscent of that of the extended version of 
SQCD studied in Section \ref{extwithout}. This analogy suggests that the 
$dP_1$ theory may lead to meta-stable minima upon the addition of extra 
massive flavors. In the next section we add massive fundamental flavors 
to the theory, and indeed find the appearance of a meta-stable SUSY breaking 
minimum.

\section{Flavored $dP_1$}
\label{flavdp1}

Inspired by the ideas presented in previous sections, we consider 
the $dP_1$ theory and explore whether the addition of light massive 
flavors for node $3$ can lead to a long-lived, meta-stable, SUSY-breaking 
minimum 
\footnote{As mentioned before, our analysis automatically generalizes to
other examples of fractional branes on obstructed geometries leading to the 
same quiver gauge theory (for instance, the fractional branes of $dP_5$ 
studied in \cite{Diaconescu:2005pc}).}. The String Theory construction 
leading to the additional fundamental flavors is provided in Section 
\ref{stringconstr}. 

\subsection{The classical flavored $dP_1$ theory}

\label{section_classical_flavored_dP1}

In the String Theory construction, see next section, we discuss that 
light massive fundamental flavors can be introduced by adding D7-branes in the 
configuration, with the new flavors arising from open strings stretched 
between D3 and D7-branes. Consistent sets of D7-branes typically add 
the same number of flavors to all gauge factors in the quiver. 

As discussed in Section \ref{stringconstr}, there are several different 
choices of a consistent set of D7-branes that can be added. These 
different choices lead in general to the same flavor content for 
the different gauge factors (i.e. the same quiver), but differ in the 
interactions of the latter with the D3-D3 states \footnote{Notice that, 
following the abuse of language of appendix B, in this discussion what we 
mean by `D3-brane' is a gauge factor in the quiver theory. Such a gauge factor can
arise from either regular or fractional D3-branes (wrapped D5-branes).}.
Specifically, different D7-branes lead to D3-D7 (and D7-D3) states with 
cubic coupling to different D3-D3 bifundamentals.

\begin{figure}[!htp]
\centering
\psfrag{3i}{$Q_{3i}$}
\psfrag{i2}{${\tilde Q}_{i2}$}
\psfrag{2j}{$Q_{2j}$}
\psfrag{j1}{${\tilde Q}_{j1}$}
\psfrag{1k}{$Q_{1k}$}
\psfrag{k3}{${\tilde Q}_{k3}$}
\includegraphics[scale=0.70]{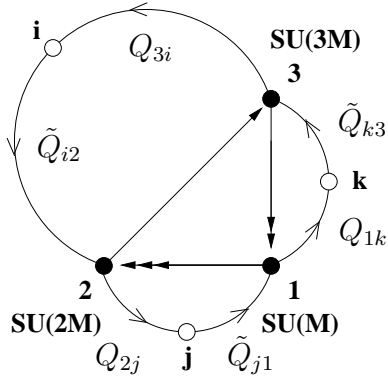}
\caption{\small
Extended quiver diagram for a $dP_1$ theory with flavors.}
\label{extd3d7}
\end{figure}

In order to keep the discussion concrete, we center on a specific set of 
D7-branes. Other choices can be analyzed similarly.
We consider three kinds of 
D7-branes, whose D3-D7, D7-D3 states couple to the 33 fields $X_{23}$, 
$X_{31}$ and $X_{12}$ respectively. In general we will consider $N_{f,1}$ 
copies of this set of D7-branes, labeled by indices $i,j,k,$ leading to 
$N_{f,1}$ additional flavors for each D3-brane gauge factor. The 
resulting gauge theory can be encoded in an {\bf extended quiver}, with 
additional nodes representing gauge symmetries on the 
D7-branes, and additional arrows representing the new flavors.
The gauge fields on the D7-brane worldvolume are higher-dimensional and 
thus appear on the four-dimensional theory as flavor global symmetries. 
The 
extended quiver diagram for this gauge theory with flavors is shown in 
Figure \ref{extd3d7}, where D7-branes are represented as white nodes. Our 
notation is that $Q_{ai}$, ${\tilde Q}_{ia}$ denote flavor fields 
associated with the $a^{th}$ D3-brane gauge factor and a D7-brane in the 
$i^{th}$ set. Notice that each of the indices for the three kinds of 
D7-branes can be regarded as an independent $SU(N_{f,1})$ global symmetry 
group. However, mass terms to be introduced later will break this 
symmetry, in general to a diagonal combination.

In addition to the superpotential (\ref{supod3}), we have a superpotential 
for the new flavors
\beqa
W_{flav.}\, =\,\lambda'\, (\,  Q_{3i} {\tilde Q}_{i2} X_{23} \, + \,
Q_{2j} {\tilde Q}_{j1} X_{12} \, +\, Q_{1k} {\tilde Q}_{k3} X_{31}\, )
\label{W_flavors_dP1}
\eeqa
where for simplicity we assume the same coupling $\lambda'$ for all terms.
We also introduce mass terms
\beqa
W_m \, =\,  m_3\, Q_{3i} {\tilde Q}_{k3} \delta_{ik} \, + \,
m_2 \, Q_{2j} {\tilde Q}_{i2} \delta_{ji} \, + \,
m_1 \, Q_{1k} {\tilde Q}_{j1} \delta_{kj} 
\eeqa
Although we work with independent masses, the general results are valid in 
the simpler situation of equal masses. Note that since the mass terms mix 
the global symmetries of the different D7-branes, we stop using 
different indexes for them.

We would like to introduce a number $N_{f,1}$ of flavors such that the 
$SU(3M)$ node is in the free magnetic phase. This corresponds to 
$N_c+1 \leq N_f < {3\over 2} N_c$. Since $N_f=N_{f,0}+N_{f,1}$ with 
$N_{f,0}=2M$, we require
\beq
M+1 \leq N_{f,1} < {5 \over 2} M
\eeq
A simple choice which works for all $M$, including $M=1$, is $N_{f,1}=2M$. 
For the moment we keep $M$ and $N_{f,1}$ general.

Let us perform a Seiberg duality transformation on node 3. The dual gauge factor 
is $SU(N)$ with $N=N_{f,1}-M$, and dynamical scale $\Lambda$. To get the 
matter content in the dual, we replace the microscopic flavors $Q_{3i}$, 
${\tilde Q}_{k3}$, $X_{23}$, $X_{31}$, $Y_{31}$ by the dual flavors 
${\tilde Q}_{i3}$, $Q_{3k}$, $X_{32}$, $X_{13}$, $Y_{13}$. We also 
have the mesons, related to the fields in the electric theory by

\beq
\begin{array}{rlcrl}
M_{21} & =\, {1\over \Lambda} X_{23} X_{31} & \quad ; \quad  & N_{k1} & 
= \, {1\over \Lambda} {\tilde Q}_{k3}X_{31} \\
M_{21}' & =\, {1\over \Lambda} X_{23} Y_{31} & \quad ; \quad  & N_{k1}' & = \, 
{1 \over \Lambda} {\tilde Q}_{k3}Y_{31} \\
N_{2i} & =\, {1\over \Lambda} X_{23} Q_{3i} & \quad ; \quad  & \Phi_{ki} & = 
\, {1\over \Lambda} {\tilde Q}_{k3}Q_{3i}
\end{array}
\eeq

There is a cubic superpotential coupling the mesons and the dual flavors
\beqa
W_{mes.} & = & h\, (\, M_{21} X_{13} X_{32} \, + \, M_{21}' Y_{13} X_{32} 
\, +\,
N_{2i} {\tilde Q}_{i3} X_{32} \, + \nonumber \\
& + & N_{k1}X_{13}Q_{3k} \, + \, N_{k1}' Y_{13} Q_{3k} \, +\, \Phi_{ki}
{\tilde Q}_{i3} Q_{3k}\, )
\eeqa
where we have taken a common coupling $h\, =\, \Lambda/\hat\Lambda$, with 
$\hat\Lambda$ related to $\Lambda$, $\Lambda_3$ by the analog of 
(\ref{match}).

In addition we have the classical superpotential, written in terms of the 
new fields
\beqa
W_{clas.} & = & h\mu_0\,(\, M_{21} Y_{12} \, -\, M_{21}' X_{12} \,)\, +\, 
\mu'\, Q_{1k} N_{k1}\, +\, \mu'\, N_{2i} {\tilde Q}_{i2} \, + \nonumber \\
& - & h\mu^{\, 2} \Tr\Phi \, 
+\, \lambda' \, Q_{2j} {\tilde Q}_{j1} X_{12} \, 
+\, m_2 Q_{2i}{\tilde Q}_{i2} \, +\, m_1 Q_{1i}{\tilde Q}_{i1}
\eeqa
where $\mu_0=\lambda \Lambda$, $\mu'=\lambda'\Lambda$, and
$\mu^{\,2}=-m_3\hat\Lambda$. Although not manifest in our 
present notation, some of the fields in this theory are close analogs of 
fields in the extended SQCD model in Section \ref{extension}. We clarify 
this analogy in Appendix \ref{section_meta_dP1}.

Some of the fields are massive, so we will proceed to integrating them 
out. However recall that $M_{21}$, $M_{21}'$, $X_{12}$, $Y_{12}$ play a 
crucial role in the dynamics of the un-flavored $dP_1$ theory. 
In fact, they are the analogs of $\Phi_{00}$ and $\Sigma_0$ in the extension of SQCD with 
massless and massive flavors. Hence it is convenient to keep them in 
the effective action until the last stage of the analysis. Thus we 
integrate out ${\tilde Q}_{i2}$, $N_{2i}$, $Q_{1k}$, $N_{k1}$.

\begin{figure}[!htp]
\centering
\psfrag{3i}{$Q_{3i}$}
\psfrag{i3}{${\tilde Q}_{i3}$}
\psfrag{i1}{${\tilde Q}_{i1}, N_{i1}'$}
\psfrag{2i}{$Q_{2i}$}
\psfrag{phi}{$\Phi$}
\includegraphics[scale=0.70]{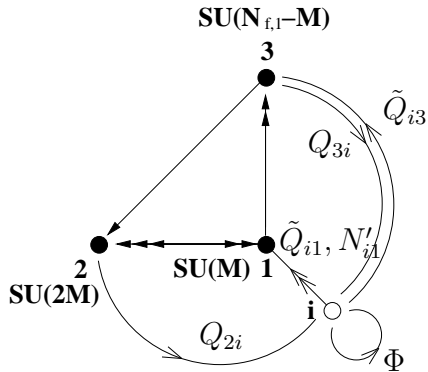}
\caption{\small
Quiver diagram for the $dP_1$ theory with flavors after dualization. 
Notice that the number of colors for nodes 1,2 and 3 are  
$M$, $2M$ and $N_{f,1}-M$. Since D7-branes are mixed after dualization, we 
represent them with a single white circle.}
\label{quiverdual}
\end{figure}

The resulting superpotential is
\beqa
W & = & h\, \Phi_{ki} {\tilde Q}_{i3} Q_{3k}\, -\, h \mu^{\, 2} \tr\Phi 
\, +\, h\mu_0\, (\, M_{21} Y_{12} \, -\, M_{21}' X_{12} \,) + \nonumber\\
& + & h\, (\, M_{21} X_{13} X_{32} \, + \, M_{21}' Y_{13} X_{32} \,
+\, N_{k1}' Y_{13} Q_{3k} \, )\, + \nonumber \\
& + & \lambda' \, Q_{2j} {\tilde Q}_{j1} X_{12} \, - \,
h_1\, {\tilde Q}_{k1}X_{13}Q_{3k} \, - \,
h_2\, Q_{2i}{\tilde Q}_{i3} X_{32} 
\label{W_dual_dP1}
\eeqa
where $h_1=m_1/\mu'$, $h_2=m_2/\mu'$. This is the theory we want to 
study. A depiction of its quiver diagram is shown in Figure 
\ref{quiverdual}.

\subsection{The local minimum}

\label{section_pmoduli_dP1}

A detailed analysis of moduli and pseudomoduli in this theory and the 
computation of pseudomoduli masses is given in Appendix \ref{section_meta_dP1}.
We present the results here. We first focus on the most symmetric choice 
of couplings $h=\lambda'=h_1=h_2$ and $\mu=\mu_0$. A discussion for less 
symetric couplings can be found in Appendix \ref{section_meta_dP1}. The 
moduli space of SUSY breaking minima is parametrized as follows

\beq
\begin{array}{c}
\begin{array}{ccccc}
{\tilde Q}_{i3}\, =\, \pmatrix{{\tilde \varphi}_1 \cr 0} & \ \ \ \ & Q_{3i}\, =\, (\varphi_1; 0) & \ \ \ \ & \Phi\, =\, \pmatrix{0 & 0 \cr 0 & \Phi_1} 
\end{array} \\ \\
\begin{array}{ccccccc}
{\tilde Q}_{k1}\, =\, \pmatrix{0 \cr y} & & N_{k1}'\, =\, \pmatrix{0 \cr z} & & Q_{2i}\, =\, \pmatrix{0 & x \cr 0 & x'} & & M_{12}'\, =\, \pmatrix{xy \cr  x'y} 
\end{array}
\end{array}
\eeq
where $\varphi_1$, ${\tilde \varphi}_1$ are $N\times N$ matrices subject to 
${\tilde \varphi}_1\varphi_1=\mu^2 \ \id_N$. The expectation values of 
all other fields vanish.

We now focus on the point of maximal unbroken global symmetry 
$\Phi_1=x=x'=y=z=(\theta+\theta^*)=0$. Computing the one loop effective 
potential we find that this critical point is a meta-stable minimum. 
Expanding  around it, e.g. for the prototypical case of $N_{f,1}=2M$, 
we get

\beq
\langle V_{eff}^{(1)}\rangle=const.+|h^4 \mu^2| \ {(\log 4-1)\over 16 \pi^2} M^2 \left(2 \ |\delta \Phi_1|^2+|\delta x|^2+|\delta x'|^2+|\delta y|^2+|\delta z|^2+|\mu^2| (\theta+\theta^*)^2\right)+\ldots
\eeq

The striking similarity between these results and those of the extended model in Section \ref{extension}
is explained in Appendix \ref{section_meta_dP1}. The longevity of this minimum is studied in Section \ref{section_longevity_dP1}.

Strictly speaking, the expectation value of $Z_{12}$ is another
pseudomodulus. This field does not appear in the classical superpotential so it is flat both at tree level and one-loop. Contrary to what happens for $X_0$ in the model of Section \ref{flavoursqcd}, motion along this direction does not take us closer to the SUSY vacua so we do not consider it poses an obvious danger. Nevertheless, it is in principle still possible that higher order corrections might render this field unstable, even modifying some of our conclusions. This is a very interesting direction for further research. In Section \ref{section_duality_cascade} we comment on another field, the
saxion, with similar behavior.

An important final comment is that the existence of SUSY breaking local 
minima depends on some basic patterns of the theory. For instance, as 
explained in Section \ref{toiss}, the fact 
that in the original theory we have $N_{f,0}<N_c$, or the existence of 
singlets (of the strongly coupled gauge factor) which couple to the 
flavors. These features are present in general quiver gauge theories of 
fractional branes at obstructed geometries, namely DSB fractional branes 
(in fact, it is easy to identify these features in all the examples of DSB 
fractional branes in \cite{Franco:2005zu}). We therefore expect that our 
conclusions for the $dP_1$ theory are of general validity for this 
whole class.

\subsection{Behavior at large fields}

\label{section_dP1_large_fields}

We now explore whether, and if so where, this theory has a SUSY vacuum.
As in previous theories, we expect to recover in the region of large fields 
the behavior of the theory without the extra flavors, determined 
in Section \ref{rundp1}, namely a runaway for $X_{12}$, $Y_{12}$.


In fact, for generic non-zero vevs of fields $\Phi$, $\tilde{Q}_{k1}$, $Q_{2i}$, $N_{k1}'$, $M_{21}$ 
and $M_{21}'$ the flavors of gauge factor 3 are massive and can 
be integrated out, leaving a pure $SU(N)$ SYM which triggers a 
non-perturbative superpotential. 
In more detail, there are a number of 
fields that contribute to the mass matrix of these flavors. Organizing 
the different fundamental flavors in a row vector $q$ with entries 
$(Q_{3k}; X_{32})$ and the anti-fundamental flavors in a column vector 
${\tilde q}=({\tilde Q}_{i3}; X_{13}, Y_{13})$, the mass matrix is given by 
\beqa
m\, =\, \pmatrix{h\Phi_{ki} & -h_1 {\tilde Q}_{k1} & hN_{k1}' \cr
-h_2 Q_{2i} & h M_{21} & h M_{21}'}
\eeqa

This matrix can be used to integrate out massive fields.
The low-energy effective dynamics is pure $SU(M)$ SYM with a 
dynamical scale $\Lambda'$ obtained from matching
\beqa
\Lambda'^{\,3M}\, =\, \frac{\det m}{\Lambda^{5M-2N_{f,1}}}
\eeqa
where $\Lambda$ is the Landau pole scale of the IR free $SU(N_{f,1}-M)$ 
theory with $2M+N_{f,1}$ flavors.

The $SU(N_{f,1}-M)$ strong dynamics generates a non-perturbative 
superpotential. The 
complete superpotential, after integrating out the massive flavors and 
taking into account the non-perturbative dynamics, becomes
\beqa
W & = & -h\mu^{\, 2}\, \tr\Phi \, +\, h\mu_0\, (\, M_{21} Y_{12} \, -\, 
M_{21}' X_{12} \,) + \,
\lambda' Q_{2j} {\tilde Q}_{j1} X_{12} \, +\, (N_{f,1}-M)\, 
\left(\, \frac{\det m}{\Lambda^{5M-2N_{f,1}}}\,\right)^{\frac 1{N_{f,1}-M}}
\nonumber
\eeqa
where one should recall that $\det m$ is a complicated function of the 
other fields in the theory.

Actually, it is easy to identify a particular direction in field space 
where the dynamics reduces to a runaway exactly like that of the 
un-flavored theory (\ref{rundp1}). Consider the equations of motion for 
$Q_{2j}$, ${\tilde Q}_{i1}$, $N_{k1}'$. It is straightforward to see that they can be 
satisfied by choosing $Q_{2j}=0$, ${\tilde Q}_{i1}=0$, $N_{k1}'=0$. Along 
this direction we have
\beqa
\det m \, =\, h^{N_{f,1}+2M}\, \det \Phi \, \det{\cal M}
\eeqa
with ${\cal M}=\pmatrix{M_{21}; M_{21}'}$. So the 
effective theory along this solution is described by the superpotential
\beqa
W & = & -h\mu^{\,2}\,  \tr\Phi \, +\, h\mu_0\, (\, M_{21} Y_{12} \, -\, 
M_{21}' X_{12} \,)\,  + \,  
(N_{f,1}-M)\, \left(\, \frac{h^{N_{f,1}+2M}\, \det \Phi \, \det{\cal 
M}}{\Lambda^{5M-2N_{f,1}}}\,
\right)^{\frac 1{N_{f,1}-M}}\nonumber
\eeqa

Using the equation of motion for $\Phi$, we obtain

\beqa
W & = & h\mu_0\, (\, M_{21} Y_{12} \, -\, M_{21}' X_{12} \,)\, 
-\, M \, \left(\, \frac{\mu^{\, 2N_{f,1}}\,
\Lambda^{5M-2N_{f,1}}}{h^{2M}\, \det{\cal M}}
\right)^{\frac 1M}
\label{W_runaway_dP1}
\eeqa
This is essentially identical to (\ref{unfldp1}), which described the 
dynamics of the un-flavored $dP_1$ theory.

\subsection{Lifetime of meta-stable vacua}

\label{section_longevity_dP1}

Comparing \eref{runsqcd} and \eref{W_runaway_dP1} and their derivations, 
we conclude that the discussion of the potential barrier height and lifetime of 
the local minimum in the flavored $dP_1$ theory is completly isomorphic 
to the one for the extension of SQCD with massless flavors of Section 
\ref{extension}.

We translate from the SQCD model to
$dP_1$ by identifying $q_1$, $\tilde{q}_1$ and $\Sigma_0$ with
$Q_{3i}$, $\tilde{Q}_{i3}$ and ${\cal Y}=\pmatrix{Y_{12}\cr X_{12}}$, 
respectively. The numbers of flavors and colors are $N_{f,0}=N_{f,1}=2M$ 
and $N=M$. Finally, $\mu$ and $\mu_0$ play the same role in both 
theories. Replacing in \eref{bounce_S_extended_model}, we get the bounce action

\beq
S={1\over |h|^{26/3}} {|\alpha_\mu|^8 \over |\epsilon|^4}
\eeq
which is independent of $M$. The interpretation of this result is identical
to the one in Section \ref{extension} and we conclude that the 
meta-stable vacua can be made parametrically long-lived.

\section{String theory construction}
\label{stringconstr}

As already mentioned, the natural way to introduce fundamental flavors in 
D3-brane quiver gauge theories is by adding D7-branes passing through the 
singular points \cite{Karch:2002sh}. This introduces a new sector of open 
strings, stretching 
between the D3 and the D7-branes, leading to such flavors. In addition, 
there is a sector of open strings stretching among the D7-branes, but the 
corresponding fields have higher-dimensional support, and they behave as 
external parameters from the viewpoint of the 4d gauge theory. In fact, 
due to the existence of superpotential couplings $X_{77'}{\tilde Q_{7'3}} 
Q_{37}$, they behave as masses for some of the flavors. Hence, D7-branes 
are the natural setup to introduce massive flavors in the string 
realization of quiver gauge theories.

To construct these configurations, we need efficient tools to classify 
interesting possibilities of D7-branes wrapped on non-compact 4-cycles on 
toric singularities, and to compute the open string 3-7 spectrum, and its 
interactions with the 3-3 sector. This study is carried out in Appendix 
\ref{D7branes} for a general toric singularity, and is applied in 
particular to the case of D7-branes in $dP_1$.

In general it is not consistent to introduce just one kind of D7-brane in 
the configuration. D7-branes of the kind constructed in Appendix 
\ref{D7branes} carry non-trivial charge under RR 4-form fields localized 
at the singularity (obtained from higher-dimensional RR $p$-forms 
integrated over the compact homology cycles of the singularity), 
hence cancellation of such RR tadpoles requires 
combinations of such branes to be introduced simultaneously. Equivalently, 
the chiral spectrum of the 4d theory obtained from just one D7-brane has 
non-abelian anomalies. Hence, only combinations of D7-branes leading to an 
anomaly-free spectrum are allowed \footnote{
The equivalence of the statements is well-known for orbifolds 
\cite{Leigh:1998hj,Aldazabal:1999nu,Bianchi:2000mi}. It 
can be argued in general as follows: Given a set of D7-branes, consider 
the net number of fundamentals minus antifundamentals they introduce for a 
given D3-brane gauge group. This number corresponds to the charge of the 
D7-brane system under the RR field associated with the compact homology 
class corresponding to that D3-brane gauge factor. Since the homology 
classes of D3-brane gauge factors form a basis of the compact homology, 
anomaly cancellation 
is equivalent to zero compact homology charge for the D7-brane system.}

\begin{figure}[ht]
  \epsfxsize = 9cm
  \centerline{\epsfbox{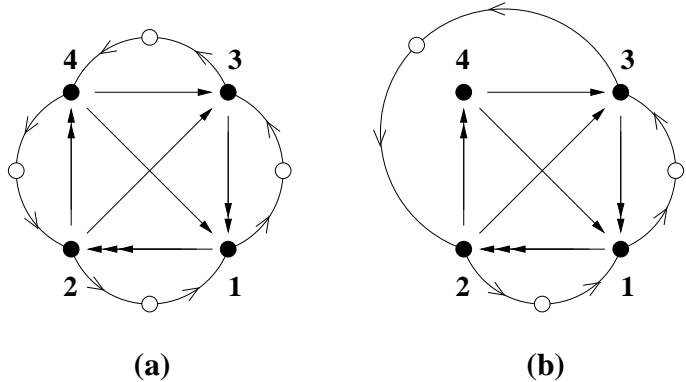}}
  \caption{Two possible extended quivers corresponding to consistent sets of 
D7-branes in the $dP_1$ theory.}
  \label{two_flavors_dP1.eps}
\end{figure}

For the case of the $dP_1$ theory, we can use the different D7-branes 
described in Appendix \ref{d7fordp1} to obtain several ways to achieve 
this. Let us consider two simple classes of solutions (although others are 
possible as well): a) the four nodes in the original quiver
get one fundamental flavor or b) only the three nodes in the final 
quiver get one fundamental flavor \footnote{Here by original and final quivers
we refer to the quivers for only regular D3-branes or only fractional branes 
respectively.}. The different possibilities of D7-branes to achieve 
this kind of spectrum are

\beq
\begin{array}{l}
\begin{array}{rl}{\bf a)} \ \ & (\Sigma_{BC}'' \, \,{\rm or}\,\, 
 \Sigma_{AB} \, \,{\rm or}\,\, \Sigma_{CD})  \oplus \ (\Sigma_{DB} 
\, \,{\rm or}\,\,  \Sigma_{CA}) \ \oplus \ \Sigma_{AD} \ \oplus \ 
(\Sigma_{DB}' \, \,{\rm or}\,\,  \Sigma_{CA}')
\end{array} \\ \\

\begin{array}{rl}
{\bf b)} \ \ & (\Sigma_{BC}'' \, \,{\rm or}\,\, \Sigma_{AB}
\, \,{\rm or}\,\, \Sigma_{CD})  \oplus \ (\Sigma_{DB}
\, \,{\rm or}\,\, \Sigma_{CA} ) \ \oplus \ \Sigma_{BC} 
\end{array}
\end{array}
\label{possible_D7s}
\eeq
where $\oplus$ denotes superposition, and `or' denotes different 
alternative possibilities. The two classes are schematically shown in 
Figure \fref{two_flavors_dP1.eps}. The white circles represent the D7-branes.

The above different possibilities lead, at the level of the quiver for 
the fractional branes, to the same spectrum of flavors, but differ in the 
superpotential couplings involving the latter. 

For the sake of concreteness we will concentrate on one particular 
example, leading to the gauge theory studied in previous section, others 
can be worked out similarly. The configuration has $2M$ copies of the 
following set of D7-branes in class $b$, leading to $2M$ additional 
flavors for nodes $1$, $2$ and $3$
\beq
\Sigma_{AB} \oplus \ \Sigma_{DB} \ \oplus \ \Sigma_{BC} \ 
\label{d7choice}
\eeq

The 33-37-73 terms in the superpotential are 
\beqa
{\tilde Q}_{i2}\, X_{23}\, Q_{3i} \quad ;
\quad {\tilde Q}_{j1}\, X_{12}\, Q_{2j} \quad ; \quad
{\tilde Q}_{k3}\, X_{31}\, Q_{1k} 
\eeqa
where $Q$, ${\tilde Q}$ correspond to 37 and 73 states, respectively, 
namely fundamental and antifundamental flavors. As mentioned, this 
configuration of D7-branes reproduces the gauge theory studied in Section 
\ref{flavdp1}, with the 4d matter content in Figure \ref{extd3d7} 
and the interactions in (\ref{W_flavors_dP1}).

A last point that requires further discussion is the introduction of 
flavor masses. As already mentioned, they are controlled by vevs for 
fields in the D7-brane sector. Namely, the configuration contains 
couplings of the form 77'-7'3-37, where $7$ and $7'$ denote different 
D7-branes simultaneously present in the configuration. Since 77' fields 
have higher-dimensional support, 
their vevs are not dynamical fields from the 4d gauge theory viewpoint, 
but rather external parameters. Hence, as in several other familiar 
situations, the moduli space of the higher-dimensional theory  
provides the parameter space of the lower-dimensional one.

The higher-dimensional theory in this case has a complicated structure, 
since it involves gauge fields in the 77 open sector, hence propagating 
over an 8d space, and charged multiplets from 77' open strings, in general 
propagating over 6d intersections. Without entering into a detailed general
discussion, it suffices our purposes to consider one particular flat 
direction of this kind of theories. Namely, we consider the mesonic flat 
direction where a set of fields $\Phi_{7_17_2}$, $\Phi_{7_27_3}$, $\ldots$, 
$\Phi_{7_k7_1}$ acquire the same vev. From the viewpoint of the 4d gauge 
theory, this implies equal mass terms for all the $37_i$, $7_i3$ fields 
coupling to them.

To be more specific, let us consider our above example, namely the choice 
of D7-branes given in (\ref{d7choice}). From the discussion in Appendix 
\ref{D7branes}, such 77'sectors exist for pairs of 
D7-branes with a common letter in their label. Thus, we see that in our 
example we have a 77' open string bi-fundamental for each pair of 
D7-branes. The above mentioned mesonic flat direction corresponds to 
recombining the three intersecting cycles into a single smooth one, which 
is at a finite distance (controlled by the D7-brane field vevs) from the 
D3-branes at the singularity. This geometrical process can be modeled as 
follows. The initial configuration contains D7-branes on three holomorphic 
4-cycles which intersect over a common holomorphic curve. In suitable 
local complex coordinates $z_1$, $z_2$, $z_3$, the 4-cycles can be chosen 
to be $z_1=0$, $z_2=0$, $z_1+z_2=0$, with the holomorphic curve thus given 
by $z_1=z_2=0$ and spanned by $z_3$. The complete D7-brane configuration 
is described by the equation $z_1z_2(z_1+z_2)=0$. Then the above mentioned 
mesonic branch corresponds to $z_1 z_2 (z_1+z_2)=\epsilon$. The branes 
have recombined since the 4-cycle is now irreducible, and all D3-D7 open
strings are massive because the 4-cycle does not pass through the origin.

\section{Embedding into a duality cascade and breaking of baryonic $U(1)$}

\label{section_duality_cascade}

So far, we have studied in detail the 3+1 dimensional gauge theory that arises 
on fractional D3-branes on a toric singularity {\it in the absence of 
regular D3-branes}. We have added fundamental flavors by means of D7-branes. 
As we have argued, it is completely licit to study theories without regular 
D3-branes. On the other hand, these theories can be regarded as describing 
the IR bottom of a duality cascade \cite{Klebanov:2000hb}. In other words,
embedding the theory in a duality cascade provides a specific UV 
completion. We now comment about a subtle point 
that should be contemplated in this case.

The gauge theory for a set of D3-branes and fractional branes at a 
singularity is not conformal and has a non-trivial RG flow. If the number 
of D3-branes $N$ is much larger than the number of fractional branes $M$, 
the theory can be regarded as a small perturbation of the
conformal theory. The general behavior is that, in analogy with the 
conifold \cite{Klebanov:2000hb}, the theory undergoes cascades of Seiberg 
dualities along which the effective number of D3-branes is 
reduced as one moves to the IR. This fact is mapped to a radial 
dependence of the 5-form flux in the corresponding gravity dual. For 
explicit examples of cascades, see 
\cite{Klebanov:2000hb,Franco:2003ja,Franco:2004jz,Franco:2005fd,Herzog:2004tr}. 
Interestingly, when a small number of D7-branes is added, the effective 
number of fractional branes is also reduced along the RG flow (see for 
example \cite{Ouyang:2003df}). This behavior 
translates into a radial 
dependence of the 3-form flux in the gravity dual \cite{Ouyang:2003df}. It is straightforward to explicitly construct 
the duality cascade that appears when regular D3-branes are added to the flavored $dP_1$ theory that we have 
focused on in this paper. Since we are really interested in the bottom IR of it, we will skip doing so. 

\begin{figure}[ht]
  \epsfxsize = 11cm
  \centerline{\epsfbox{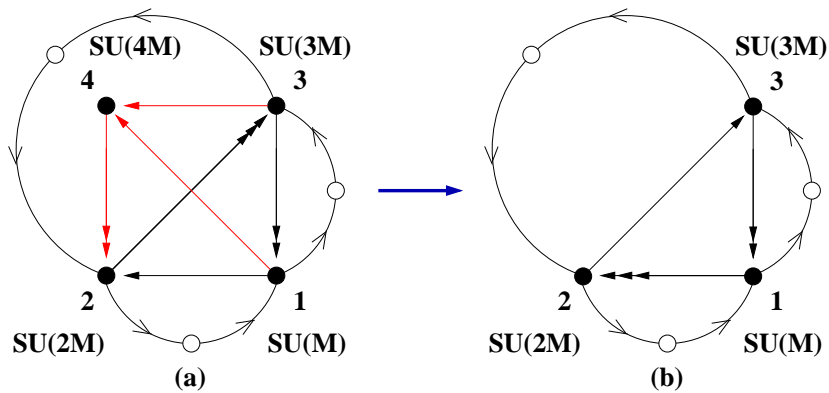}}
  \caption{a) Quiver diagram for flavored $dP_1$ with $M$ regular D3-branes and $M$ D5-branes. b) The theory on the baryonic branch.}
  \label{dP1_quiver_last}
\end{figure}

The number of D3, D5 and D7-branes in the UV can be appropriately chosen such that after a large number 
of dualizations we reach a point where the $N=M$ as shown in \fref{dP1_quiver_last} \footnote{When doing a Seiberg duality 
transformation on a node, fundamental flavors of other nodes can appear as Seiberg mesons combining bifundamental 
and (anti)fundamental fields as in Section \ref{section_classical_flavored_dP1}. Because of this, it seems possible to have a cascade in which 
the quiver in \fref{dP1_quiver_last}.a is periodically repeated up to a change in the numbers of D3 and D5-branes and possibly 
permutations of the nodes.}.

This theory, without the fundamental flavors, has been already investigated 
in \cite{Bertolini:2005di}, to which we refer the reader for details. We 
are interested in the situation in which the dynamics of node $4$ becomes 
dominant. Since our choice of D7-branes is such that node $4$ has no 
fundamental flavors, the discussion in \cite{Bertolini:2005di} applies 
without changes. Since node $4$ has an equal number of colors and flavors 
it has a quantum modified moduli space that is realized by adding the 
constraint $\det \mathcal{M}-B \tilde{B}=\Lambda_4^{8M}$ to the superpotential 
via a Lagrange multiplier. The mesons are combinations of the fields that 
we have indicated in red in \fref{dP1_quiver_last}. 
Thorough analysis shows that the mesonic branch is completely lifted in 
this theory. Along the baryonic branch, we obtain the flavored $dP_1$ 
theory of Section \ref{flavdp1} (shown in \fref{dP1_quiver_last}.b).

But we have to be cautious at this point. The global $U(1)$ baryonic 
symmetry of present in the original theory is spontaneously broken by the vevs of dibaryonic 
operators. As a result, the IR theory also contains a massless pseudo-scalar Goldstone boson (the "axion"). By 
supersymmetry, the axion falls into a massless $\mathcal{N}=1$ chiral multiplet. Then, there will also
be a massless scalar (the "saxion") and a Weyl fermion (the "axino") . The axion and the saxion are combined
into the complex scalar of the chiral multiplet. The above argument is the
generalization of the analysis in \cite{Gubser:2004qj} of the conifold 
cascade.

While the axion is a Goldstone boson and remains massless, the flatness 
of the saxion can in principle be lifted by quantum corrections. At 
one-loop the saxion is decoupled from the flavored $dP_1$ sector and the 
computations of previous sections are not modified. Namely, the saxion 
remains massless at one-loop. It is not clear 
whether the saxion becomes unstable at higher loops or whether it can 
change any of our conclusions by coupling to the flavored $dP_1$ fields. 
We think that this is an interesting problem, important for the
possible realization of gravitational duals of our theories, and which 
hence deserves further study.

\section{Conclusions}

In this paper we have studied the generalization of the ISS proposal to 
diverse gauge theories with massless flavors, including quiver gauge 
theories on fractional branes. Interestingly enough, the requirements 
of the ISS proposal (like SUSY breaking by the rank condition mechanism) 
suggest that the natural generalization for quiver gauge 
theories occurs for fractional branes in geometries with obstructed complex 
deformations. Although our detailed analysis has centered on concrete 
examples, our results have laid the grounds for more general analysis of 
this class of models. 

Thus, it would be interesting to extend our computations to
arbitrary number of extra flavors and of colors / fractional branes.
Also, it would be interesting to analyze the introduction of flavors in 
other simple examples of DSB fractional branes (like the $dP_2$ or 
$dP_3$ theories).

The generalization of the ISS proposal for fractional branes in  
obstructed geometries leads effectively to a mechanism that 
allows to stay away from the runaway behavior of these configurations. 
This is an important development, that improves the possible application 
of these theories to dynamical SUSY breaking in String Theory model 
building (see \cite{Diaconescu:2005pc} for a partial attempt). 

It would be interesting to extend our discussion to other quiver gauge theories, 
for instance those arising in the presence of orientifold planes, so as to exploit 
the ISS mechanism for SO and Sp gauge theories. If the appropriate conditions are 
met, it is possible that addition of flavors can be used to escape the runaway 
behavior of models like that in \cite{Lykken:1998ec}.

Despite the progress made in this paper, we consider that it is important to be 
cautious about the fate of fields that remain decoupled from the rest of the 
theory and flat up to one-loop, such as $Z_{12}$ in Section 
\ref{section_pmoduli_dP1}. This is an issue that needs to be understood in more 
detail. In particular it is worthy to understand whether they can become unstable 
and, if so, whether they modify any of our conclusions.

Another important open question concerns the realization of gravitational 
duals of the gauge theories at SUSY breaking minima. The generalization of 
the ISS mechanism to quiver gauge theories carried out in this paper is an 
important step in this direction. However, several other questions 
remain open. One of them is the saxion flat direction mentioned in
Section \ref{section_duality_cascade}. Another important point is that
the large number of flavor branes requires the construction of
supergravity solutions including their backreaction, which are
very involved even in simple examples
(for some discussions see e.g.
\cite{Bertolini:2001qa,Ouyang:2003df,Burrington:2004id,Casero:2006pt}).

We expect the fascinating physics of dynamical supersymmetry breaking 
and its realization in String Theory to continue triggering progress in 
these and other directions.

\section*{Acknowledgments}

We thank M. Bertolini, I. Garc\'{\i}a-Etxebarria, I. Klebanov, M. 
Kruczenski, L. McAllister, C. Nu\~nez, F. Saad, M. Wijnholt for useful 
discussions. We are specially grateful to K. Intriligator for reading
the manuscript and useful comments. We thank the Department of Physics and 
Astronomy at UPenn, USA, and Centro At\'omico Bariloche, Argentina, for 
hospitality during completion of this work. A.U. thanks M. Gonz\'alez for 
encouragement and support.


\appendix

\section{Computation of pseudomoduli masses}

\label{comput}

The one-loop correction to the vacuum energy due to integrating out classically massive
fields is 
\beq
V^{(1)}_{eff} = {1\over
 64\pi^2}{\rm STr}\,\CM^4\log{\CM^2\over\Lambda ^2}\equiv
 {1\over 64\pi ^2}\left( \Tr \, m_B^4 \log {m_B^2\over \Lambda ^2}
 -\Tr \, m_F^4 \log {m_F^2\over \Lambda ^2}\right)
\label{V_1_loop}
\eeq
where $m_B^2$ and $m_F^2$ are the classical squared masses for bosons
and fermions as functions of the pseudomoduli vevs\footnote{The ultraviolet cutoff
$\Lambda$ can be absorbed in the renormalization of the couplings that appear in
the tree-level vacuum energy.}.

In a theory of $n$ chiral superfields $Q^a$ with canonical
classical K\"ahler potential, $K_{cal}=Q_a^\dagger Q^a$ and superpotential $W(Q_a)$, the scalar
and fermion mass-squared matrices are given by
\beq
m_0^2=\pmatrix{W^{\dagger ac}W_{cb}&W^{\dagger abc}W_c\cr
W_{abc}W^{\dagger c}&W_{ac}W^{\dagger cb}} \qquad
m_{1/2}^2=\pmatrix{W^{\dagger ac}W_{cb}&0\cr 0&W_{ac}W^{\dagger
cb}}
\label{m_scalars_fermions}
\eeq
where $W_c\equiv \partial W/\partial Q^c$, etc. The dimension of $m_0^2$
and $m_{1/2}^2$ is $2n\times 2n$. Supersymmetry breaking is encoded in the off-diagonal
blocks of these matrices.

Typically, the effective potential for the pseudomoduli \eref{V_1_loop} generates 
masses for pseudomoduli when expanded around its critical points. These 
masses can be positive (the corresponding pseudomodulus is a stable direction) or negative
(unstable direction). Sometimes, pseudomoduli remain massless at one-loop. Since they are not
Goldstone bosons, their masses are not protected from perturbative corrections and it is
therefore expected that their flatness is lifted at some higher order.

\subsection{Massless flavored SQCD}

\label{section_meta_toy}

The superpotential for this theory is
\beqa
W & = & h\, 
\Tr q_0 \Phi_{00} {\tilde q}_0\, + \, \Tr q_0 \Phi_{01} {\tilde q}_1\, + \, 
\Tr q_1 \Phi_{10} {\tilde q}_0\, + \, \Tr q_1 \Phi_{11} {\tilde q}_1\, 
- \, h \mu^2 \Tr \Phi_{11}
\eeqa

The F-term for $\Phi_{11}$ breaks supersymmetry due to the rank 
condition. The classical minima of this potential are obtained by 
saturating this F-term as much as possible. There is a moduli space of field 
configurations satisfying this with $V_{min}=(N_{f,1}-N)|h^2 \mu^4|$. In particular, for any choice of vevs for 
$q_1$, ${\tilde q}_1$ of the form
\beqa
q_1 \, =\, \pmatrix{\, \varphi_1\, ;\, 0\,} \quad  \quad
{\tilde q}_1 \, =\, \pmatrix{\, {\tilde \varphi}_1\, \cr \, 0\,}
\label{pm_massless_SQCD_1}
\eeqa
with ${\tilde \varphi}_1\varphi_1=\mu^2 \id_N$. Here ${\tilde \varphi}_1$, $\varphi_1$ are $N\times N$ blocks.

In order to make the F-terms of $\Phi_{01}$, $\Phi_{10}$ vanish, $q_0$ and ${\tilde q}_0$ must vanish. 
In addition, to make the F-terms of $q_1$ and ${\tilde q}_1$ vanish, the vevs of $\Phi_{10}$, 
$\Phi_{01}$, $\Phi_{11}$ must be of the form
\beqa
\Phi_{01}\, =\, \pmatrix{ \, 0 \, ;\, Y\, } \quad  \quad
\Phi_{10}\, = \, \pmatrix{\, 0\, \cr \, {\tilde Y} \,}  \quad  \quad
\Phi_{11}\, =\, \pmatrix{\, 0\,\,\, 0 \, \cr \, 0 \, X_1 \, }
\label{pm_massless_SQCD_2}
\eeqa
where $Y$, ${\tilde Y}$, $X_1$ are $N_{f,0}\times (N_{f,1}-N)$,
$(N_{f,1}-N)\times N_{f,0}$ and $(N_{f,1}-N)\times (N_{f,1}-N)$ blocks, 
respectively. Finally, the vev for $\Phi_{00}=X_0$ is arbitrary.

We can use $SU(N_{f,0})$ global symmetry transformations to make $X_0$ 
diagonal. Furthermore, $X_1$ can be diagonalized by means of $SU(N-N_{f,1})$ 
transformations.

We now expand fields in fluctuations around arbitrary expectation values of the 
the pseudomoduli
\beq
\begin{array}{c}
q_0  = \delta \rho_0 \ \ \ \ \  {\tilde q}_0 \, =\, \delta{\tilde \rho}_0 \ \ \ \ \  
q_1  = \, \pmatrix{\, \mu e^{\theta} \id_N \, + \, \delta \chi\,\, ; \,\, \delta\rho_1\,} \ \ \ \ \  
{\tilde q}_1 \, = \, \pmatrix{\, \mu e^{-\theta} \id_N \, + \, \delta {\tilde \chi}\, \cr \, \delta{\tilde \rho}_1\, } \\ \\
\begin{array}{lcl}
\Phi_{00}  = \, X_0\, +\, \delta \Phi_0 & \ \ \ \ \ & \Phi_{01} \, =\, \pmatrix{\, \delta W\, ;\, Y+\delta Y\, } \\ \\
\Phi_{1,0}\, =\, \pmatrix{\, \delta{\tilde W} \, \cr {\tilde Y} +\delta{\tilde Y}}  & \ \ \ \ \ & 
\Phi_{11}  = \pmatrix{\, \delta Y_1 \, & \, \delta Z_1 \, \cr\, \delta{\tilde Z}_1 \, & \, X_1 {\id_{N_f-N}} \, + \, \delta \Phi_1} 
\end{array}
\end{array}
\label{expansion_toy_model}
\eeq

We must now expand the classical superpotential to quadratic order in the 
fluctuations, except for the terms involving $\delta\Phi_1$, for which we 
must allow cubic terms. The reason is that $\delta\Phi_1$ is the only 
field with non-vanishing F-term in the vacuum, so it leads to a 
contribution to the scalar mass matrix involving third derivatives of the 
terms in which $\delta\Phi_1$ appears. The result is
\beqa
W & = & h\, \Tr \left[\delta\rho_1 \, (\, X_1\, +\, \delta\Phi_1\, )\, \delta{\tilde \rho}_1
\, -\, \mu^2 \, (\, X_1 + \delta\Phi_1\, ) \, 
+\, \mu e^\theta \delta Z_1 \delta{\tilde \rho}_1 \, + \,
\mu e^{-\theta} \delta \rho_1 \delta{\tilde Z}_1 \,
+ \right.\nonumber \\
& + & \mu e^\theta \delta Y_1 \delta{\tilde \chi}_1 \, +\,
\mu e^{-\theta} \delta \chi_1 \delta{\tilde Y}_1 \, +\, 
\mu e^{-\theta} \delta \rho_0 \delta W \, +\,
\mu e^{\theta} \delta {\tilde W} \delta{\tilde \rho}_0 \, +  \nonumber \\
& + & \left.  \delta\rho_0 \, 
X_0\, \delta{\tilde \rho}_0 \, +\,
\delta\rho_0 \, Y \delta{\tilde \rho}_1 \, + \, \delta\rho_1 \, 
{\tilde Y}\, \delta{\tilde \rho}_0 \right]
\label{expansion_toy}
\eeqa

As in SQCD with massive flavors \cite{Intriligator:2006dd}, off diagonal 
elements of $\delta\Phi_1$ do not enter in the mass matrix. The same 
thing happens for $\delta\Phi_0$. This is natural since they are Goldstone 
bosons of the broken global symmetries, which obviously remain 
exactly massless (to any order).

The fields $\delta Y_1$, $\delta\chi_1$, $\delta{\tilde \chi}_1$ are decoupled 
from the SUSY breaking sector (i.e. from fields with a non-supersymmetric
mass matrix). As a result, they have a supersymmetric mass matrix and do 
not contribute to the supertrace.

The main difference with respect to the case studied in 
\cite{Intriligator:2006dd} is given by the generically rectangular  
matrices $Y$ and $\tilde{Y}$ coupling fluctuations. Nevertheless, in
analogy with \cite{Intriligator:2006dd}, one can use symmetries of the 
system to show that the lagrangian of pseumoduli masses is given by a sum 
of terms $\tr M^\dagger M$, where $M$ denotes the different matrix 
pseudomoduli (see (\ref{m1}) later). The coefficients of such terms can 
be computed by taking a simple ansatz for the corresponding 
pseudomodulus. That is, we take $Y$ to be formed by a diagonal block 
$diag(Y_1,\ldots,Y_{\tilde{N}})$, where 
$\tilde{N}=\min(N_{f,0},N_{f,1}-N)$, and an appropriately located block of 
zeroes (i.e. either additional rows or columns) to complete
the $N_{f,0} \times (N_{f,1}-N)$ dimensions. We take the matrix 
$\tilde{Y}$ to have an analogous block-diagonal form.

Now, the interpretation of \eref{expansion_toy} is clear, relating it to 
O'Raighfertaigh-like (OR) models of the kind used in  
\cite{Intriligator:2006dd}. 
For $(N_{f,1}-N) < N_{f,0}$ it corresponds to the sum of $(N_{f,1}-N)$ copies 
of an OR model involving all terms in \eref{expansion_toy}. For $N_{f,0} 
< (N_{f,1}-N)$ we have $N_{f,0}$ copies of an OR model with all the terms
in \eref{expansion_toy}, plus $(N_{f,1}-N-N_{f,0})$ copies of an OR model 
in which all the terms in \eref{expansion_toy} except the last two are 
present (this latter OR model effectively corresponds to that used in 
\cite{Intriligator:2006dd}). 

We then compute $m_0^2$ and $m_{1/2}^2$ using \eref{m_scalars_fermions} 
and plug the result into \eref{V_1_loop} to obtain the one-loop effective 
potential. The full expression for $V^{(1)}_{eff}$ is very complicated. 
For illustrative purposes we present it for fixed values 
$Y=\tilde{Y}=\theta=0$

{\footnotesize
\beq
\begin{array}{rl}
V^{(1)}_{eff}|_{Y,\tilde{Y},\theta=0}(X_1,X_0)=& {h^4 N(N_f-N) \over 128 \pi^2} \left[4 (\mu^2+X_1^2)^2 \log \left({ h^2 (\mu^2+X_1^2) \over \Lambda }\right) \right. \\
+ & \left(3 \mu^2+X_1^2-\sqrt{\mu^4+6\mu^2 X_1^2+X_1^4}\right)^2 \log \left( {h^2 \left( 3 \mu^2+X_1^2-\sqrt{\mu^4+6\mu^2 X_1^2+X_1^4}\right) \over 2 \Lambda }\right) \\
+ & \left(3 \mu^2+X_1^2+\sqrt{\mu^4+6\mu^2 X_1^2+X_1^4}\right)^2 \log \left( {h^2 \left( 3 \mu^2+X_1^2+\sqrt{\mu^4+6\mu^2 X_1^2+X_1^4}\right) \over 2 \Lambda }\right) \\
+ & 2 \left(2 \mu^2+X_1 \left(X_1 -\sqrt{4 \mu^2+X_1^2}\right)\right)^2 \log \left( {h^2 \left(2 \mu^2+X_1 \left(X_1 -\sqrt{4 \mu^2+X_1^2}\right)\right) \over 2 \Lambda }\right) \\
- & 2 \left(2 \mu^2+X_1 \left(X_1 +\sqrt{4 \mu^2+X_1^2}\right)\right)^2 \log \left( {h^2 \left(2 \mu^2+X_1 \left(X_1 +\sqrt{4 \mu^2+X_1^2}\right)\right) \over 2 \Lambda }\right)
\end{array}
\eeq}
which does not depend on $X_0$ at all. In order to keep the above expression compact, we have omitted absolute values. 

The full one-loop effective potential has a critical point at the vacua 
of maximal unbroken global symmetry, which correspond to 
$X_0=X_1=Y=\tilde{Y}=0$ and $\varphi_1={\tilde \varphi}_1=\mu$ (up to 
unbroken flavor rotations).

Expanding the effective potential around these pseudomoduli vevs, we obtain 

\beq
\langle V_{eff}^{(1)}\rangle=const.+|h^4 \mu^2| \ {(\log 4-1)\over 16 \pi^2}N \left[(N_f-N)\left(2 \ |\delta \Phi_1|^2+|\mu^2| (\theta+\theta^*)^2\right)+|\delta Y|^2+|\delta \tilde{Y}|^2 \right]+\ldots
\label{m1}
\eeq
where $|\delta Y|^2$ and $|\delta \tilde{Y}|^2$ should be understood as the squared norms of $\tilde{N}$-dimensional complex vectors. Taking the non-trivial
diagonal blocks of $Y$ and $\tilde{Y}$ to be proportional to $\id_{\tilde{N}}$, \eref{m1} simplifies to 

\beq
\langle V_{eff}^{(1)}\rangle=const.+|h^4 \mu^2| \ {(\log 4-1)\over 16 \pi^2}N\left[(N_f-N)\left(2 \ |\delta \Phi_1|^2+|\mu^2| (\theta+\theta^*)^2\right)+\tilde{N}\left(|\delta Y|^2+|\delta \tilde{Y}|^2 \right)\right]+\ldots
\label{m1_2}
\eeq

\begin{figure}[ht]
  \epsfxsize = 8cm
  \centerline{\epsfbox{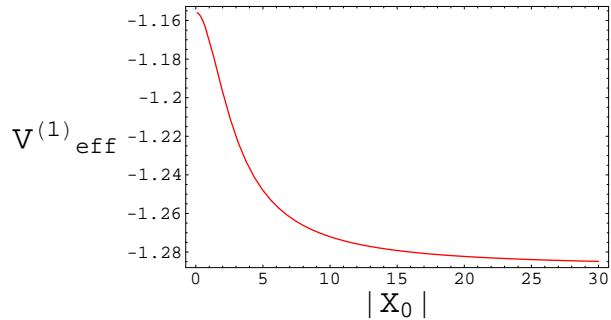}}
  \caption{The one-loop effective potential in arbitrary units as a function of $\Phi_0$ for $Y \neq 0$ and $\Phi_1=\tilde{Y}=\theta=0$.}
  \label{X0_unstable}
\end{figure}

From \ref{m1}, we see that $\delta \Phi_0$ remains massless at one-loop, i.e. $X_0$ remains a flat direction. An heuristic, although not rigorous, way to understand why this happens is to
look at \eref{expansion_toy} and notice that for $Y=\tilde{Y}=0$, $X_0$ decouples from the SUSY breaking sector and thus
it does not contribute to the supertrace, disappearing from $V^{(1)}_{eff}$. 
Outside the critical point, for $Y,\tilde{Y}\neq 0$, $X_0$ couples to the SUSY breaking sector and appears in the effective potential, 
but it becomes an unstable direction as shown in \fref{X0_unstable}.

In principle, it is still possible that $\delta \Phi_0$ becomes massive at higher loops, producing a meta-stable minimum (probably with a 
much smaller potential barrier) at small expectation values for the 
fields. We do not consider this possibility but explore a different direction in Section \ref{extension}. The {\bf most economical} way to lift this flat direction classically is by considering a modified toy model,
with the addition to the electric theory of a neutral field $\Sigma_0$ with cubic coupling to massless flavors.

\subsection{Extended model}

\label{section_meta_extended}

The pseudomoduli in this case are identical to those in \eref{pm_massless_SQCD_1}
and \eref{pm_massless_SQCD_2} with the exception that the coupling \eref{extsupo} 
`freezes' the expectation value of $X_0$ to zero in the magnetic theory as discussed 
in Section \ref{extension}. This occurs already at the classical level so $X_0$ is not a pseudomoduli 
in this theory.

Repeating the derivation in the previous section, the expansion of the superpotential
to second order in fluctuations (and to third order in terms involving $\delta \Phi_1$ which
has a non-vanishing F-term) is

\beqa
W & = & h\, \Tr \left[\delta\rho_1 \, (\, X_1\, +\, \delta\Phi_1\, )\, \delta{\tilde \rho}_1
\, -\, \mu^2 \, (\, X_1 + \delta\Phi_1\, ) \, 
+\, \mu e^\theta \delta Z_1 \delta{\tilde \rho}_1 \, + \,
\mu e^{-\theta} \delta \rho_1 \delta{\tilde Z}_1 \,
+ \right.\nonumber \\
& + & \left. \mu e^{-\theta} \delta \rho_0 \delta W \, +\,
\mu e^{\theta} \delta {\tilde W} \delta{\tilde \rho}_0 \, +\,
\delta\rho_0 \, Y \delta{\tilde \rho}_1 \, + \, \delta\rho_1 \, 
{\tilde Y}\, \delta{\tilde \rho}_0 \right]
\label{expansion_extended}
\eeqa
where we have already dropped fields that do not couple to the SUSY 
breaking sector. The effective potential has a critical point at the 
vacua of maximal unbroken global symmetry given, up to unbroken
flavor rotations, by $X_1=Y=\tilde{Y}=0$ and $\varphi_1={\tilde 
\varphi}_1=\mu$. 

As in the previous section, we can split the lagrangian for the fluctuations
into a sum of simple OR modes.

Expanding $V^{(1)}_{eff}$ around these vacua we obtain

\beq
\langle V_{eff}^{(1)}\rangle=const.+|h^4 \mu^2| \ {(\log 4-1)\over 16 \pi^2}N \left[(N_f-N)\left(2 \ |\delta \Phi_1|^2+|\mu^2| (\theta+\theta^*)^2\right)+|\delta Y|^2+|\delta \tilde{Y}|^2 \right]+\ldots
\label{m2}
\eeq
i.e. all pseudomoduli are lifted at one-loop and we have a SUSY breaking, meta-stable minimum
at $X_1=Y=\tilde{Y}=0$ and $\varphi_1={\tilde \varphi}_1=\mu$. Once again, taking the
diagonal blocks of $Y$ and $\tilde{Y}$ to be proportional to $\id_{\tilde{N}}$, \eref{m2} reduces to 

\beq
\langle V_{eff}^{(1)}\rangle=const.+|h^4 \mu^2| \ {(\log 4-1)\over 16 \pi^2}N\left[(N_f-N)\left(2 \ |\delta \Phi_1|^2+|\mu^2| (\theta+\theta^*)^2\right)+\tilde{N}\left(|\delta Y|^2+|\delta \tilde{Y}|^2 \right)\right]+\ldots
\label{m2_2}
\eeq

\subsection{Flavored $dP_1$}

\label{section_meta_dP1}

The quiver diagram for this model is shown in \fref{quiverdual}. The superpotential
is given by \eref{W_dual_dP1} and corresponds to

\beqa
W & = & h\, \Phi_{ki} {\tilde Q}_{i3} Q_{3k}\, -\, h \mu^{\, 2} \tr\Phi 
\, +\, h\mu_0\, (\, M_{21} Y_{12} \, -\, M_{21}' X_{12} \,) + \nonumber\\
& + & h\, (\, M_{21} X_{13} X_{32} \, + \, M_{21}' Y_{13} X_{32} \,
+\, N_{k1}' Y_{13} Q_{3k} \, )\, + \nonumber \\
& + & \lambda' \, Q_{2j} {\tilde Q}_{j1} X_{12} \, - \,
h_1\, {\tilde Q}_{k1}X_{13}Q_{3k} \, - \,
h_2\, Q_{2i}{\tilde Q}_{i3} X_{32} 
\eeqa

The first step is to parametrize the SUSY breaking vacua and to identify the pseudomoduli. 
Given the complicated structure of the theory this task is slightly more involved than in
previous examples. There are $15$ chiral fields in this theory. We simply impose the 
minimization of their F-terms (actually all F-terms, except for the one of $\Phi$, vanish) 
in a convenient order. 


We start with 
\beqa
\frac{\partial W}{\partial \Phi}= 0 \quad \to \quad {\tilde 
Q}_{i3}Q_{3k}=\mu^2 \ \id_{N_{f,1}}
\eeqa
This cannot be saturated and breaks SUSY by the rank condition. This F-term contribution 
is minimized by choosing
\beqa
{\tilde Q}_{i3}\, =\, \pmatrix{{\tilde \varphi}_1 \cr 0} \quad ; \quad
Q_{3i}\, =\, (\varphi_1; 0)
\eeqa
where $\varphi_1$, ${\tilde \varphi}_1$ are $N\times N$ matrices and 
${\tilde \varphi}_1\varphi_1=\mu^2 \ \id_N$.

Now we impose

\beqa
\partial_{Y_{12}} W & =\, 0 \quad \to \quad & M_{21}  =\, 0
\nonumber \\
\partial_{N_{k1}'} W & =\, 0 \quad \to \quad & Y_{13} Q_{3k}  
=\, 0 \quad \to \quad Y_{13}\, =\, 0 \nonumber \\
\partial_{M_{21}'} W & =\, 0 \quad \to \quad & Y_{13} X_{32} 
-\mu_0 X_{12} =\, 0 \quad \to \quad X_{12}\, =\, 0 
\label{W_dP1}
\eeqa

So far, the results are independent of the presence of couplings $h_1$ and $h_2$. 
From now on, we make explicit use of them (in their absence, vevs are less 
constrained, leading to more pseudomoduli). Let us continue by imposing

\beqa
\partial_{{\tilde Q}_{i1}} W & =\, 0 \quad \to \quad &
\lambda' X_{12}Q_{2i}\, -\, h_1 X_{13}Q_{3i} =\, 0 \quad \to \quad X_{13}\, =\, 0 \nonumber \\
\partial_{Q_{2i}} W & =\, 0 \quad \to \quad &
\lambda'{\tilde Q}_{i1} X_{12} \, - \, h_2 {\tilde Q}_{i3}X_{32} \, =\, 0
\quad \to \quad X_{32}\, =\, 0 \nonumber \\
\partial_{M_{21}} W & =\, 0 \quad \to \quad &
X_{13}X_{32}+\mu_0 Y_{12} \, =\, 0 \quad \to \quad Y_{12}\, =\, 0
\eeqa

With this, we are ready to get the expression for $\Phi$. We have

\beqa
\partial_{{\tilde Q}_{i3}} W & =\, 0 \quad \to \quad &
h Q_{3k}\Phi_{ki}\, -\, h_2 X_{32} Q_{2i} \, =\, 0
\quad \to \quad Q_{3k}\Phi_{ki}\, =\, 0 \nonumber \\
\partial_{Q_{3k}} W & =\, 0 \quad \to \quad &
h \Phi_{ki} {\tilde Q}_{i3}\, -\, h_1 {\tilde Q}_{k1} X_{13} \, + N_{k1}' 
Y_{13}\, =\, 0 \quad \to \quad
\Phi_{ki} {\tilde Q}_{i3}\, =\, 0 
\eeqa

From these two conditions $\Phi$ has the structure
\beqa
\Phi\, =\, \pmatrix{0 & 0 \cr 0 & \Phi_1}
\eeqa
Notice that the fact that $h_1$ and $h_2$ are non-vanishing is crucial for 
getting this. Continuing, we have 

\beqa
\partial_{X_{13}} W & =\, 0 \quad \to \quad &
h X_{32}M_{21} \, - \, h_1 Q_{3k} {\tilde Q_{k1}}=\, 0
\quad \to \quad {\tilde Q}_{k1}\, =\, \pmatrix{0 \cr y} \nonumber\\
\partial_{Y_{13}} W & =\, 0 \quad \to \quad &
X_{32}M_{21}' \, + \,  Q_{3k} N_{k1}'=\, 0
\quad \to \quad N_{k1}'\, =\, \pmatrix{0 \cr z} \nonumber\\
\partial_{X_{32}} W & =\, 0 \quad \to \quad &
h M_{21}X_{13} \, + \, h M_{21}' Y_{13} \, - \,  h_2 Q_{2i} {\tilde Q}_{i3}
\, =\, 0 \quad \to \quad Q_{2i}\, =\, \pmatrix{0 & x \cr 0 & x'} 
\nonumber \\
\partial_{X_{12}} W & =\, 0 \quad \to \quad &
-h \mu_0 M_{21}'\, + \, \lambda' Q_{2j}{\tilde Q}_{j1} \, =\, 0 \quad \to \quad
M_{12}'\, =\, {\lambda' \over h \mu_0}\pmatrix{xy \cr  x'y} 
\eeqa

So far we have minimized $14$ of the F-term contributions. We are only left with
$Z_{12}$ which does not appear in the superpotential in \eref{W_dP1} and thus its contribution
trivially vanishes.

Notice that only the vev of $M_{12}'$ depends on superpotential couplings. 
Furthermore, none of the vevs depend on $h_1$ or $h_2$.
These facts will be important when studying the effective potential for general
values of the couplings.

Expanding the superpotential to quadratic order in fluctuations, except for terms involving $\delta \Phi_{11}$ (the only field with a non-zero F-term in the vacuum) which have to be expanded to cubic order, we obtain

{\footnotesize
\beq
\begin{array}{rl}
W= & \Tr \left[- h \, \mu^2 \Phi_1 + h \, \Phi_1 \delta Q_{3,2} \delta \tilde{Q}_{3,2}+ h \, \mu \, e^{\theta} \delta Y_{13} \delta N_{1,1}-h_1 \, \mu \, e^{\theta} \delta X_{13} \delta \tilde{Q}_{1,1}-h_1 \, y \, \delta X_{13} \delta Q_{3,2}+h \, z \, \delta Y_{13} \delta Q_{3,2} \right. \\
+& {x \, y \, \lambda' \over \mu_0} \delta Y_{13} \delta X_{32,1}- h_2 \, \mu  \, e^{-\theta} \delta Q_{2,11}\delta X_{32,1}-h_2 \, x \, \delta \tilde{Q}_{3,2}\delta X_{32,1}+{x' y \, \lambda' \over \mu_0} \delta Y_{13} \delta X_{32,2}- h_2 \, \mu \, e^{-\theta} \delta Q_{2,21} \delta X_{32,2} \\ 
-& \left. h_2 \, x' \, \delta \tilde{Q}_{3,2} \delta X_{32,2}+h \, \mu \, e^{\theta} \delta \tilde{Q}_{3,2} \delta \Phi_{01}+h \, \mu \, e^{-\theta} \delta Q_{3,2} \delta \Phi_{10}-h \, \mu^2 \delta \Phi_{11}+h \, \delta Q_{3,2}  \delta \tilde{Q}_{3,2} \delta \Phi_{11} \right] 
\end{array}
\label{expansion_W_dP1}
\eeq} where we have only kept those terms that couple to supersymmetry
breaking fields and thus give a non-vanishing contribution to the
supertrace when integrated out. We have used the obvious notation for
fluctuations, including additional subindices to indicate matrix
sub-blocks (that reduce to matrix entries for $M=1$). 
In analogy with \eref{expansion_toy_model}, $\theta$ is defined by

\beqa
{\tilde Q}_{i3}\, =\, \pmatrix{\mu e^{\theta} {\id_N} +\delta {\tilde 
Q}_{3,1} \cr {\tilde Q}_{3,2}} \quad ; \quad
Q_{3i}\, =\, (\mu e^{-\theta} {\id_N}+\delta Q_{3,1}; \delta Q_{3,2})
\eeqa

Let us consider in a little more detail those fluctuations that are absent
from \eref{expansion_W_dP1} because they have a supersymmetric
matrix. Fluctuations of classically massive fields $\delta M_{21}$,
$\delta M_{21}'$, $\delta X_{12}$ and $\delta Y_{12}$ are naturally
expected not to contribute to the effective potential. Also $\delta
Q_{3,1}$, $\delta \tilde{Q}_{3,1}$ and $\delta \Phi_{11}$ do not
contribute, in complete analogy with similar fields in SUSY QCD with
massive flavors.  The only new ingredient is the fact that $\delta
Q_{2,12}$ and $\delta Q_{2,22}$ disappear.

Until now, we have kept our discussion completely general. In order to 
compute the effective potential we have to diagonalize the mass matrices 
\eref{m_scalars_fermions}. Doing this analytically is intractable for 
generic values of the superpotential couplings. We now focus on the most 
symmetric choice of couplings, i.e. $h=\lambda'=h_1=h_2$ and $\mu=\mu_0$. 

As in the previous examples, the lagrangian for the fluctuations splits 
into a sum  of OR models. We now focus in the case $N_{f,1}=2M$. Since 
$N_{f,0}=2M$, this implies $N=M$.
The effective potential has a minimum at $\Phi_1=x=x'=y=z=(\theta+\theta^*)=0$. 
Expanding around it, we have  

\beq
\langle V_{eff}^{(1)}\rangle=const.+|h^4 \mu^2| \ {(\log 4-1)\over 16 \pi^2} M^2 \left(2 \ |\delta \Phi_1|^2+|\delta x|^2+|\delta x'|^2+|\delta y|^2+|\delta z|^2+|\mu^2| (\theta+\theta^*)^2\right)+\ldots
\label{m_dP1}
\eeq

This result is remarkably similar to the to the one for the 
extension of SQCD with massless flavors of Appendix \ref{section_meta_extended}. This is 
due
to the close similarity between \eref{expansion_extended} and \eref{expansion_W_dP1}. 
In fact, it is possible to identify analogous fields in both models

\beq
\begin{array}{cccl}
\mbox{{\bf Extended massless SQCD}} & \ \ \ \ \ & \mbox{{\bf Flavored $dP_1$}} & \\
\Phi_{00} & & M_{21}, M_{21}' & \\
\Phi_{01} & & N'_{i1} & \mbox{similar mesons, $N_{i1}$, become massive} \\  
\Phi_{10} & & N_{2i} & \mbox{they are massive} \\
\Phi_{11} & & \Phi &
\end{array}
\eeq \nonumber

The analogue of $\Phi_{10}$ in flavored $dP_1$ are the flavors $N_{2i}$ coming out of node $2$. These fields are massive and do not appear in the final theory. Despite this, $Q_2$ takes a vev of similar form to that of $\Phi_{10}$. Similarly, $\tilde{Q}_1$ is not analogous to $\Phi_{01}$ but its vevs have the same structure.

\subsection*{Extending the analysis to different couplings}

We have just considered a very symmetric situation in which all dimensionless couplings
in the superpotential are identical. It is natural to wonder whether meta-stable minima
still exist in the case in which the couplings are different. In order to study this question
we consider the case in which $h_1=h_2=\lambda'$, but we allow $h_1$ to be different from 
$h$. In this case we cannot treat the problem analytically anymore and we 
proceed numerically.

As before, $V_{eff}^{(1)}$ has a critical point for $\Phi_1=x=x'=y=z=(\theta+\theta^*)=0$. Expanding around it we conclude that the mass of the $x$, $x'$ and $\theta+\theta^*$ fluctuations depend only on $h$ while masses of $y$ and $z$ fluctuations seem to be equal and depend on both $h$ and $h_1$.
Near the critical point, we have\footnote{Although we have computed the effective potential numerically for specific values of $h$, $h_1$ and $\mu$, we 
provide an analytical expression in \eref{m_dP1_generic}. With our analysis, we can 
only say that this expression 
is correct to a high numerical accuracy.}

\beq
\begin{array}{rl}
\langle V_{eff}^{(1)}\rangle= & const.+|h^4 \mu^2| \ {(\log 4-1)\over 16 \pi^2}\left(2 \ |\delta \Phi_1|^2+|\delta x|^2+|\delta x'|^2+|\mu^2| (\theta+\theta^*)^2\right)+ \\ 
+ & m_y^2 \ |\delta y|^2+m_z^2 \ |\delta z|^2+\ldots
\end{array}
\label{m_dP1_generic}
\eeq

\begin{figure}[ht]
  \epsfxsize = 8cm
  \centerline{\epsfbox{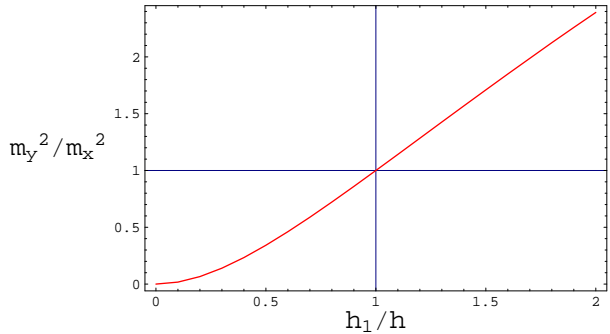}}
  \caption{The value of $m^2_y/m^2_x$ as a function of $h_1/h$. We indicate the point of symmetric couplings $h_1=h$.}
  \label{my_h1}
\end{figure}

\fref{my_h1} shows the behavior of $m^2_y=m^2_z$ as a function
of $h_1/h$.

We can consider more involved situations in which all couplings are 
different. For small variations of the couplings, the existence of a 
meta-stable minimum is guaranteed. This follows because the moduli space 
of the theory for equal couplings is compact, and the small variation of 
couplings can be regarded as a small potential on it. We have moreover 
performed a numerical analysis in some directions in coupling space.
Our analysis seems to indicate that the existence of meta-stable, SUSY 
breaking minima where all pseudomoduli get positive masses at one-loop 
and are consequently lifted is robust with respect to variations of the 
couplings.

\section{Flavor D7-branes for D3-branes at singularities from dimers}

\label{D7branes}

The introduction of flavors in gauge theories realized on systems of
D3-branes at singularities is naturally achieved by incorporating
D7-branes. Such systems have been considered in geometries related
to flat space or orbifolds in e.g.
\cite{Park:1998zh,Park:1999eb,Park:1999ep,Bertolini:2001qa,Karch:2002sh}.
However, for D3-branes at non-orbifold singularities, it is non-trivial to obtain 
the D3-D7 open string spectrum
and interactions (see e.g. \cite{Park:1999ep,Ouyang:2003df} for discussion 
in some simple examples). In this appendix we introduce techniques to 
construct and characterize a simple class of D7-branes wrapped on 
holomorphic 4-cycles in systems of D3-branes at general toric Calabi-Yau 
threefold singularities. Our characterization is based on the description 
of these systems in terms of dimer diagrams (or brane tilings) 
\cite{Hanany:2005ve,Franco:2005rj,Hanany:2005ss,Feng:2005gw,Franco:2006gc}. 
More specifically, our main tool is the Riemann surface $\Sigma$ in the 
mirror configuration, studied in \cite{Feng:2005gw} (and whose skeleton 
is the web diagram).

Before entering the discussion, a comment is in order. In this appendix 
(and pieces of the main text using its results), each gauge factor arising 
on the D3-brane gauge theory is loosely referred to as a `D3-brane'. 

\subsection{General lessons from $dP_0$}

\label{D7dp0}

\begin{figure}[!htp]
\begin{center}
\epsfxsize=5cm
\hspace*{0in}\vspace*{.2in}
\epsffile{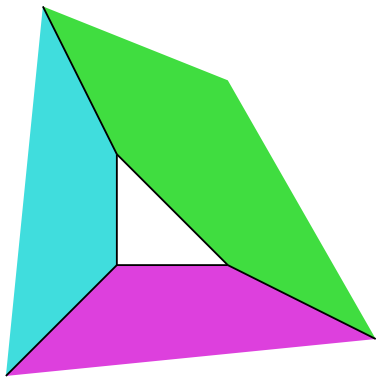}
\caption{\small
Three basic 4-cycles in the $dP_0$ theory.}
\label{cyclesweb}
\end{center}
\end{figure}

Let us start with a heuristic argument. In systems of D3-branes at 
singularities, supersymmetric flavor D7-branes wrap holomorphic non-compact 
4-cycles. A set of these (the so-called toric divisors) are associated with 
external points in the toric diagram, or non-compact faces in web 
diagrams, spanned by adjacent external legs. To be concrete, the web 
diagram of the complex cone over $dP_0$, namely the $\IC^3/\IZ_3$ orbifold 
singularity, is shown in Figure \ref{cyclesweb}, along with the three 
basic non-compact 4-cycles. They correspond to the three 4-cycles 
defined by $z_i=0$, $i=1,2,3$, where $z_i$ denote complex coordinates of 
$\IC^3$, which descend to the orbifold space.

This orbifold example already shows a crucial subtlety. Starting with 
D7-branes wrapped on a 4-cycle e.g. $z_1=0$ of the parent $\IC^3$, 
 there are three possible choices of Chan-Paton factor in quotienting by 
$\IZ_3$, given by the three roots of unity. This shows that for each 
4-cycle there are three different discrete choices that define a 
D7-brane in the orbifold geometry. In fact, different choices of 
Chan-Paton factors lead to different 
D3-D7 spectra, etc. Hence it is important to characterize this subtle 
feature already in the quotient space. By doing so we will be able to 
generalize to other examples, including non-orbifold ones, where the 
notion of Chan-Paton factor is actually not so familiar.

\begin{figure}[!htp]
\begin{center}
\epsfxsize=10cm
\hspace*{0in}\vspace*{.2in}
\epsffile{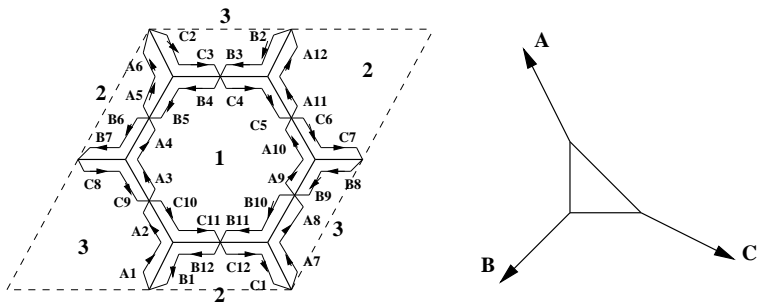}
\caption{\small
Dimer and zig-zag paths in the $dP_0$ theory.}
\label{dp0dimer}
\end{center}
\end{figure}
\begin{figure}[!htp]
\begin{center}
\epsfxsize=5cm
\hspace*{0in}\vspace*{.2in}
\epsffile{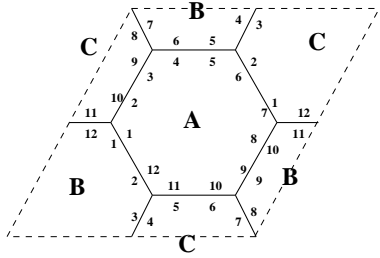}
\caption{\small
Mirror Riemann surface for the $dP_0$ theory. Sides in this picture are 
identified so that the topology of $\Sigma$ is that of a (punctured) genus 
1 Riemann surface. Punctures are located in the middle of the faces in 
the picture. The fact that the tiling of $\Sigma$ is similar to the 
original brane tiling is a property of del Pezzo theories, and not valid 
for a general singularity.} 
\label{dp0riemann}
\end{center}
\end{figure}

The structure of D7-branes, including discrete multiplicities, turns 
out to be very simple in the mirror geometry. Recall that in the 
mirror geometry a prominent role is played by the (punctured) Riemann 
surface $\Sigma$ whose skeleton is the web diagram. This Riemann surface 
can be obtained by considering the zig-zag paths of the dimer diagram, 
which provide a tiling of $\Sigma$ where faces correspond to zig-zag 
paths, and their adjacency can be read from the dimer diagram. In Figure 
\ref{dp0dimer} we give the dimer diagram and zig-zag paths for the 
$dP_0$ theory. The mirror Riemann surface $\Sigma$ is shown in Figure 
\ref{dp0riemann}.

\begin{figure}[!htp]
\begin{center}
\epsfxsize=3.5cm
\hspace*{0in}\vspace*{.2in}
\epsffile{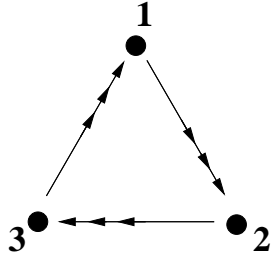}
\caption{\small
Quiver for the $dP_0$ theory.}
\label{dp0quiver}
\end{center}
\end{figure}

\begin{figure}[!htp]
\begin{center}
\epsfxsize=12cm
\hspace*{0in}\vspace*{.2in}
\epsffile{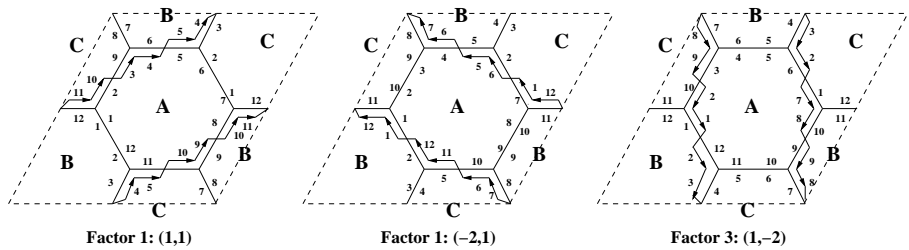}
\caption{\small
The 1-cycles for the three D3-brane gauge factors in the $dP_0$ theory.}
\label{dp0d3fin}
\end{center}
\end{figure}

\begin{figure}[!htp]
\begin{center}
\epsfxsize=4cm
\hspace*{0in}\vspace*{.2in}
\epsffile{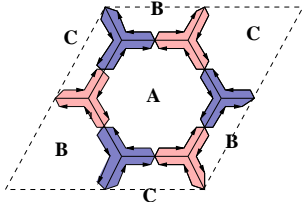}
\caption{\small
Disks corresponding to the cubic couplings in the superpotential of the 
$dP_0$ theory.} 
\label{dp0supo}
\end{center}
\end{figure}

In this mirror picture, the different D3-brane gauge factors arise from 
D6-branes on compact 3-cycles, which are encoded in non-trivial 
compact 1-cycles in $\Sigma$. Amusingly, these are zig-zag paths of the 
tiling of $\Sigma$. Moreover, the  bifundamental chiral multiplets 
arise from intersections of these 3-cycles, and  their superpotential 
couplings arise from disks bounded by the 1-cycles. The 1-cycles 
associated with the three D3-brane gauge factors are shown in Figure 
\ref{dp0d3fin}. Notice the geometric $\IZ_3$ symmetry (mirror to the 
quantum $\IZ_3$ symmetry of the orbifold) exchanging them. It is easy to 
check the intersection numbers of the 1-cycles and recover the quiver 
diagram for the $dP_0$ theory, shown in Figure \ref{dp0quiver}. Also, the 
cubic superpotential couplings can be obtained from triangles bounded by 
pieces of 1-cycles in $\Sigma$, see Figure \ref{dp0supo}.

\begin{figure}[!htp]
\begin{center}
\epsfxsize=5cm
\hspace*{0in}\vspace*{.2in}
\epsffile{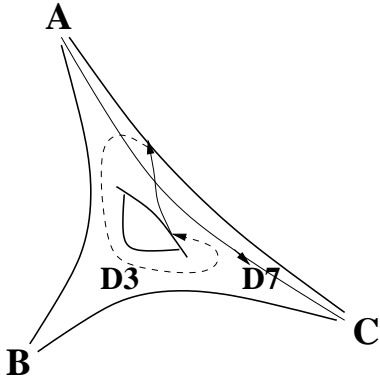}
\caption{\small
Intuitive picture of the 1-cycles in $\Sigma$ associated with D3 and 
D7-branes.} 
\label{intuiriemann}
\end{center}
\end{figure}

It is now easy to describe the D7-branes in this picture. In the mirror 
geometry, they correspond to D6-branes wrapped on non-compact 3-cycles.
These correspond to non-compact 1-cycles in $\Sigma$, which come from 
infinity at one puncture and go to infinity at another puncture 
\footnote{The complete 3-cycle, along with its supersymmetry, are 
discussed later on.} (so in 
the web diagram they are naturally associated with non-compact faces 
spanned by the corresponding legs). An intuitive picture of the 1-cycles 
in $\Sigma$ associated with D3 and D7-branes is shown in Figure 
\ref{intuiriemann}.

\begin{figure}[!htp]
\centering
\psfrag{z1}{$z_1=0$}
\psfrag{z2}{$z_2=0$}
\psfrag{z3}{$z_3=0$}
\includegraphics[scale=0.4]{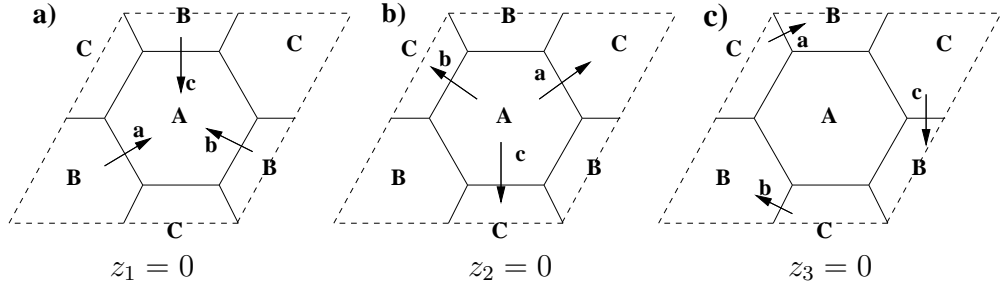}
\caption{\small
Different 1-cycles corresponding to the mirrors of D7-brane 4-cycles for 
the $dP_0$ theory. The segments represent 1-cycles stretched between the 
punctures located at the position of the face labels.}
\label{dp0d7s}
\end{figure}

The origin of the Chan-Paton multiplicity is clear in this description.
As one can see in Figure \ref{dp0riemann}, for each pair of punctures 
there are three possible paths that the associated D7-brane 1-cycle can 
take. These are explicitly shown in Figure \ref{dp0d7s}. 
For fixed pair of punctures, the different 1-cycles correspond to 
different possible supersymmetric  D7-branes wrapped on the corresponding 
4-cycle. Namely, to different choices of Chan-Paton factors. Indeed, for 
a fixed pair of punctured the different 1-cycles are related by the 
$\IZ_3$ geometric symmetry, mirror to the $\IZ_3$ quantum symmetry of the 
orbifold theory, in agreement with the Chan-Paton interpretation. 
To make contact with standard orbifold notation, 1-cycles of type BA, AC, 
CB correspond to D7-branes wrapped on the 4-cycles $z_i=0$, 
for $i=1,2,3$ respectively (often denoted D7$_1$, D7$_2$, D7$_3$ in the 
orbifold/orientifold literature). On the other hand, the choices $a$, $b$, 
$c$ correspond to different choices of the Chan-Paton action.

\begin{figure}[!htp]
\begin{center}
\epsfxsize=5cm
\hspace*{0in}\vspace*{.2in}
\epsffile{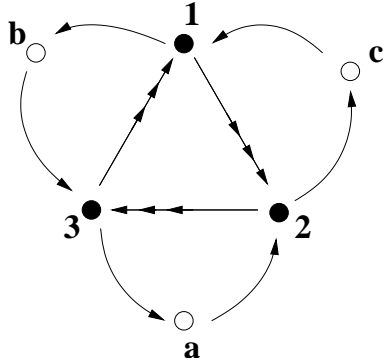}
\caption{\small
Extended quiver for the $dP_0$ theory with D7-branes.}
\label{dp0extquivers}
\end{center}
\end{figure}

It is easy to obtain the 3-7 spectrum of chiral multiplets by simply 
computing the intersection numbers of the 1-cycles associated with the 
D7-brane of interest, and the 1-cycles associated with the gauge factors of 
the D3-branes. This gives rise to the extended quivers in Figure 
\ref{dp0extquivers}. This agrees with the spectra that one can find using 
techniques of D-branes at orbifolds and Chan-Paton factors.

\begin{figure}[!htp]
\begin{center}
\epsfxsize=12cm
\hspace*{0in}\vspace*{.2in}
\epsffile{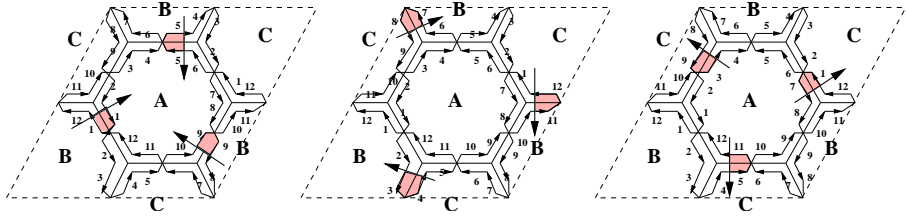}
\caption{\small
Disks corresponding to 33-37-73 interactions in the $dP_0$ theory.}
\label{d37supo}
\end{center}
\end{figure}

Notice that different D7-branes on the same 4-cycle but with different 
choice of Chan-Paton factors lead to different D3-D7 spectrum. Also
notice that different D7's can give rise to the same spectrum of 3-7 
chiral multiplets. However, different possibilities give rise to different 
superpotential interactions of type 33-37-73. These interactions are 
easily computed by considering disks bounded by two D3-branes and one
D7-brane, shown in Figure \ref{d37supo}. These are also in 
agreement with results from orbifold computations (they correspond to 
the interactions usually denoted $(33)_i$-$37_i$-$7_i3$, where $(33)_i$ 
denotes a bi-fundamental arising from the orbifold projection of the 
chiral multiplet parametrizing motion in the complex direction $z_i$ in 
the parent $\mathcal{N}=4$ theory of D3-branes in flat space).

Recall that each edge in the tiling of $\Sigma$ corresponds to a 
bi-fundamental multiplet in the 33 sector. Hence our example illustrates that 
for each such bi-fundamental there is a possible supersymmetric D7-brane, 
given by the 1-cycle crossing the edge. Moreover, a little thought 
on the above pictures reveals that a bi-fundamental 
$\Phi_{ij}=(\fund_i,\antifund_j)$ is associated with a D7-brane whose 73 
and 37 sectors transform as $\antifund_i$ and $\fund_j$ respectively, and 
that there is a 33-37-73 interaction between the three fields. These 
features follow from the dimer construction and are valid for general 
singularities.

\medskip

Finally, in the presence of several D7-branes, there are in general 
non-trivial D7-D7' open string sectors. These are higher-dimensional 
fields, but are relevant for the 4d theory due the existence of 
37-77'-7'3 couplings. Namely, vevs for 77' fields appear as mass terms for 
flavors of the D3-brane gauge theories.

The 77' fields and their interactions, can be determined using orbifold techniques. 
Concretely, two D7-branes on different 4-cycles $z_i=0$, $z_j=0$, (denoted 
D7$_i$-D7$_j$) and with different Chan-Paton actions lead to one 
six-dimensional 7$_i$-7$_j$ field (propagating on the extra $z_i=z_j=0$ complex 
plane). The dimensionality of the D7-branes intersection is determined by the
number of common punctures along which the corresponding non-compact 1-cycles go
to infinity. At the origin, one of the 4d $\mathcal{N}=1$ chiral multiplets in this 6d 
hypermultiplet has a superpotential coupling to the flavors in the 
D3-D7$_i$ and D3-D7$_j$ sectors. Also, two D7-branes on the same 4-cycle 
$z_i=0$ (denoted D7$_i$, D7$_i$') but different choices of Chan-Paton 
action lead to one eight-dimensional field. At the origin one 4d $\mathcal{N}=1$ 
chiral multiplet couples to the flavors in the D3-D7$_i$ 
and D3-D7$_i$' sectors. Finally, pairs of D7-branes with same choice of 
Chan-Paton action lead to higher-dimensional fields, but which do not couple
to D3-D7 states (the orbifold projection forces these D7-brane fields to 
vanish at the origin).

\begin{figure}[!htp]
\begin{center}
\epsfxsize=6cm
\hspace*{0in}\vspace*{.2in}
\epsffile{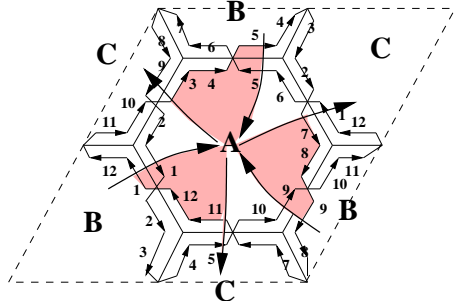}
\caption{\small
Examples of interactions $37_i-7_i7_j-7_j3$ in the $dP_0$ theory.}
\label{dp0d7supo}
\end{center}
\end{figure}

\begin{figure}[!htp]
\begin{center}
\epsfxsize=3.5cm
\hspace*{0in}\vspace*{.2in}
\epsffile{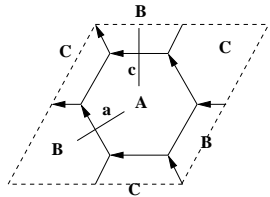}
\caption{\small
The non-trivial interaction $37_i-7_i7_i'-7_i'3$, which can be shown to 
exist using orbifold techniques, is not manifest as a 
disk in the mirror Riemann surface.}
\label{nodisk}
\end{center}
\end{figure}

In contrast with 33 and 37 states, 77' states are not perfectly 
characterized in the dimer diagram. This is related to the fact that
they have non-compact support, hence they cannot be properly described
in terms of intersections of 1-cycles (which lead to essentially 
four-dimensional fields). Heuristically, one could associate such fields 
to `intersections' of D7-brane 1-cycles with common punctures. 
Using this picture, some $37_i-7_i7_j-7_j3$ interactions can 
be pictured in terms of disks as shown in Figure \ref{dp0d7supo}. However, notice that there are 
additional interactions that cannot visualized in this way. For 
instance, the couplings $37_i-7_i7_i'-7_i'3$ exist, despite 
the fact that they do not correspond to disks in $\Sigma$ (see Figure 
\ref{nodisk}). Hence, interactions of $77'$ states with the 4d theory 
cannot be directly read out from the dimer diagram.

As mentioned, vevs for 77' states correspond to mass terms for 
certain D3-D7 open strings. Sketchily, the string interpretation of this 
fact is that the 4-cycles associated with the D7 and the D7' recombine into 
a 4-cycle which does not pass through the singular point, but at some 
distance from it. A prototypical example (e.g. in $\IC^3$) is to recombine 
D7-branes along $z_1=0$ and along $z_2=0$ (i.e. $z_1z_2=0$) to D7-branes 
along $z_1z_2=\epsilon$. This gives a non-trivial mass to the 37 and 7'3 
open strings, which have a minimal non-zero stretching related to 
$\epsilon$. 


As a final piece of information, we would like to provide a more detailed 
description of the 3-cycles in the mirror geometry corresponding to the D7-branes, and to discuss 
their supersymmetry properties. Recall that the mirror geometry is given 
as a $\Sigma \times \IC^*$ fibration over a complex plane $\IC$. Let $z$ 
be a (uniform) complex coordinate in our genus $1$ Riemann surface,  
and let the $\IC^*$ fibration be described by $xy=w$, where $w$ is the 
complex coordinate on the $\IC$ base. The holomorphic 3-form of 
the geometry is
\beqa
\Omega\, =\, dw\, dz \, \frac{dx}{x}
\eeqa
D3-branes are mirror to D6-branes on 3-cycles which span the $\IS^1$ 
direction in $\IC^*$ (given by the orbit of $x\to e^{i\, t} x$, $y\to 
e^{-i\,t} y$), times a segment in $\IC$ (locally of the form  
$w=e^{i\theta} r$), times a 1-cycle in $\Sigma$ (locally of the form 
$z=e^{-i\theta} s$. Here $r,s,t$ are local coordinates on the 3-cycle.
The 3-cycle is supersymmetric since it is special lagrangian with respect 
to $e^{-i\pi/2}\Omega$ (namely they are calibrated by ${\rm Im}\,\Omega$). 
Indeed 
\beqa
\Omega|_{D3}\, =\, i dr \, ds\, dt
\eeqa
So ${\rm Re}\, \Omega|_{D3}=0$, ${\rm Im}\,\Omega|_{D3}=d{\vol}_3$.

The 3-cycles mirror to D7-branes correspond to 1-cycles parallel to some 
D3-brane 1-cycle in $\Sigma$ (compare Figures \ref{dp0d3fin} and 
\ref{dp0d7s}), hence along $z=e^{-i\theta}$ (note however 
that they stretch between punctures, so they are non-compact). In 
addition, we need to specify the two additional directions. They span 
a semi-infinite line in $\IC$ described by $w=e^{i\theta} s$ for $s\ge 
0$, and the $\IS^1$ direction in $\IC^*$. Such 3-cycles are non-compact 
and calibrated by  ${\rm Im}\,\Omega$, hence preserve the same 
supersymmetry as the D3-brane 3-cycles
\footnote{Clearly, there are other non-compact and supersymmetric 
3-cycles in the geometry (mirror to other B-type branes, like D9-branes 
with holomorphic gauge bundles). The identification of the above ones as mirror 
of the D7-branes is ensured by the fact that the intersection numbers 
with the D3-brane 3-cycles reproduce the same D3-D7 spectrum as with 
orbifold techniques.}.

A last important point is that for the above 3-cycles, the intersection 
numbers of the different 3-cycles is simply given by the intersection 
numbers of the 1-cycles in $\Sigma$.

\subsection{Generalization}

\label{generatoric}

The above story admits a natural generalization to any toric singularity.
The general lesson is that D7-branes wrapped on holomorphic 4-cycles 
correspond to 1-cycles stretching between two punctures in the Riemann 
surface \footnote{\label{baryons} This relation between bi-fundamentals 
and 4-cycles is isomorphic to another familiar relation, see 
e.g.\cite{Franco:2005sm}: given a bi-fundamental multiplet, one 
can consider the corresponding dibaryonic operator. In the dual 
AdS$_5\times Y_5$, where $Y_5$ is the base of the conical singularity $X_6$, 
the dual to the dibaryon is a D3-brane wrapped on a supersymmetric 
3-cycle $C_3$ in $X_5$. The cone over $C_3$ is a holomorphic 4-cycle in 
$X_6$. This is the 4-cycle which we are using, in a different setup, to 
wrap our D7-branes. Notice that the discussion below on Chan-Paton 
factors is isomorphic to that in section 2.2.1 of \cite{Franco:2005sm}.
Also note that this correspondence between bi-fundamentals and D7-branes 
is valid for a general singularity, even non-toric ones.}. 
More specifically, for each bi-fundamental field in the D3-D3 
sector, one can construct a supersymmetric D7-brane, with a 
superpotential coupling 33-37-73 to precisely such bi-fundamental 
\footnote{It is important to clarify that in general all these D7-branes 
are not independent. In the language of footnote \ref{baryons}, the 
dibaryons of the bi-fundamental fields are not all independent. Rather, we 
use this rule to generate a (possibly redundant) class of D7-branes, 
their D3-D7 spectra and their interactions, in a simple fashion.}.

As discussed above, this rule is manifest in the dimer graph. Moreover it 
implies that for a fixed choice of a pair of punctures, i.e. for a 
fixed 4-cycle, there may be a multiplicity of different D7-branes, which 
differ by the choice of `Chan-Paton action'. For non-orbifold singularities, 
this requires some explaining. Even in non-orbifold examples, D7-branes 
carry world-volume gauge degrees of freedom. Topologically different 
choices correspond to different D7-branes, in the sense that their D3-D7 
spectrum and interactions are different. For instance, the choice of 
Chan-Paton factors in an orbifold model corresponds to such a topological 
choice, given by the holonomy of the gauge connection at infinity. This 
notion is however general, and can be used even in non-orbifold examples. 
Namely, the region at infinity in a non-compact 4-cycle is given, for 
toric geometries, by a Lens space $\IS^3/\IZ_n$. The value of $n$ for a 
given 4-cycle can be obtained from the web diagram (equivalently from the 
1-cycle in $\Sigma$) by computing the bilinear form $n=p_1q_2-q_1p_2$ for 
the $(p_i,q_i)$ charges of the legs/punctures associated with it. Since 
$H_1(\IS^3/\IZ_n)=\IZ_n$, the holonomy of the gauge connection at 
infinity is characterized by an element of $\IZ_n$. This implies that 
there are $n$ possible choices of asymptotic behavior of the Chan-Paton 
bundle corresponding to D7-branes on such 4-cycle. Effectively, this 
corresponds to $n$ different choices of D7-brane, in the sense that each 
choice leads to a different D3-D7 spectrum and interactions. 

In fact this is in remarkable agreement with the dimer diagram picture, 
where there are indeed $n$ different ways to connect the two punctures by 
crossing an edge in the tiling of $\Sigma$.

Given a D7-brane, the computation of the D7-D3 spectrum is given by the 
intersection numbers of the corresponding 1-cycles, as described above. 
Similarly, the 37-73-33 interactions correspond to disk diagrams in 
$\Sigma$, and lead to a coupling of the D3-D7 branes to the 33 
bi-fundamental associated to the D7-brane.

Finally, the 77' sector and its interactions with D3-D7 states are not 
properly encoded in the dimer diagram. We leave the general question of 
characterizing this sector for non-orbifold singularities as an open 
question. In the next section, the results we use for the $dP_1$ theory 
are obtained by requiring consistency upon higgsing the $dP_1$ to the 
$dP_0$ theory. For a general toric singularity, the interactions between 
77' and D3-D7 states can be determined by computing them in a sufficiently large
abelian orbifold and partially resolving it to obtain the singularity of interest.

\subsection{D7-branes for the $dP_1$ theory}
\label{d7fordp1}

\begin{figure}[!htp]
\begin{center}
\epsfxsize=10cm
\hspace*{0in}\vspace*{.2in}
\epsffile{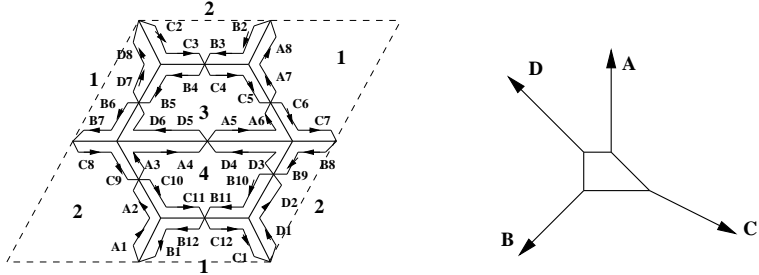}
\caption{\small
(a) Dimer diagram and zig-zag paths in the $dP_1$ theory. (b) Web diagram.}
\label{dp1dimer}
\end{center}
\end{figure}

Let us now consider the case of interest in the main text, namely the 
$dP_1$ theory. The unit cell of the dimer diagram and the zig-zag paths 
are shown in Figure \ref{dp1dimer}a. The web diagram is shown in Figure 
\ref{dp1dimer}b.

\begin{figure}[!htp]
\begin{center}
\epsfxsize=6cm
\hspace*{0in}\vspace*{.2in}
\epsffile{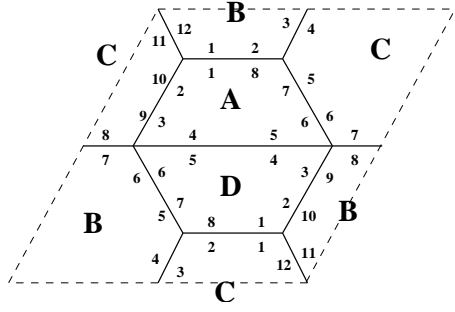}
\caption{\small
Mirror Riemann surface for the $dP_1$ theory. The fact that it comes out 
to be similar to the original brane tiling is a property of del Pezzo 
theories, and not valid for a general singularity.} 
\label{dp1riemann}
\end{center}
\end{figure}

The mirror Riemann surface is shown in Figure \ref{dp1riemann}.

Notice the close relation with the $dP_0$ theory, which amounts (up to a 
trivial relabeling) to the removal of the edge separating faces 3 and 4 in 
the dimer diagram, and of the edge separating faces A and D in the tiling 
of $\Sigma$.

\begin{figure}[!htp]
\begin{center}
\epsfxsize=10cm
\hspace*{0in}\vspace*{.2in}
\epsffile{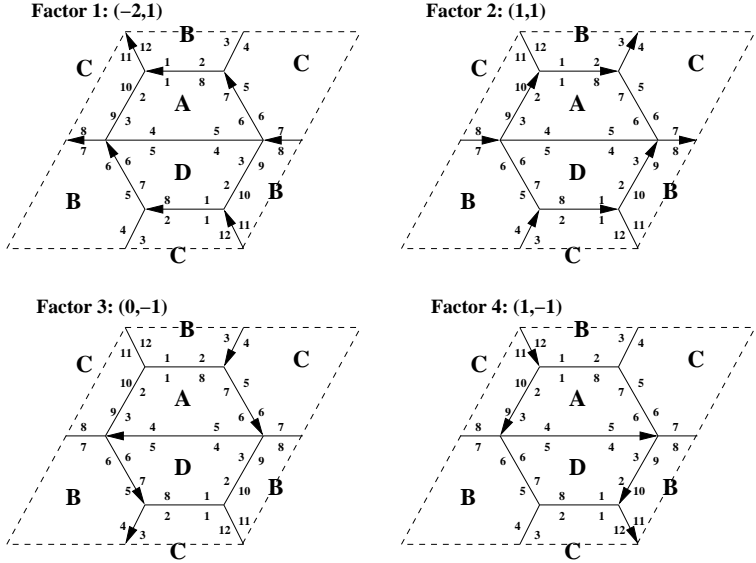}
\caption{\small
The 1-cycles for the four D3-brane gauge factors in the $dP_1$ 
theory. }
\label{dp1d3s}
\end{center}
\end{figure}

The 1-cycles in $\Sigma$ corresponding to the different D3-brane gauge 
group factors are shown in Figure \ref{dp1d3s}

\begin{figure}[!htp]
\begin{center}
\epsfxsize=3cm
\hspace*{0in}\vspace*{.2in}
\epsffile{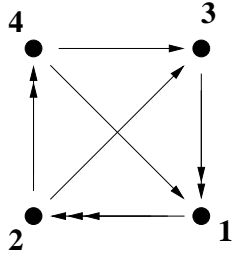}
\caption{\small
Quiver diagram for the $dP_1$ theory.} 
\label{dp1quiver}
\end{center}
\end{figure}

Their intersection number reproduces the quiver of the $dP_1$ theory, see 
Figure \ref{dp1quiver}

\begin{figure}[!htp]
\centering
\psfrag{ab}{$\Sigma_{AB}$}
\psfrag{ca1}{$\Sigma_{CA}$}
\psfrag{ca2}{$\Sigma_{CA}'$}
\psfrag{ad}{$\Sigma_{AD}$}
\psfrag{bc1}{$\Sigma_{BC}$}
\psfrag{bc2}{$\Sigma_{BC}'$}
\psfrag{bc3}{$\Sigma_{BC}''$}
\psfrag{db1}{$\Sigma_{DB}$}
\psfrag{db2}{$\Sigma_{DB}'$}
\psfrag{cd}{$\Sigma_{CD}$}
\psfrag{x31}{$X_{31}$}
\psfrag{y31}{$Y_{31}$}
\psfrag{x24}{$X_{24}$}
\psfrag{y24}{$Y_{24}$}
\psfrag{x12}{$X_{12}$}
\psfrag{y12}{$Y_{12}$}
\psfrag{z12}{$Z_{12}$}
\psfrag{x43}{$X_{43}$}
\psfrag{x23}{$X_{23}$}
\psfrag{x41}{$X_{41}$}
\includegraphics[scale=0.4]{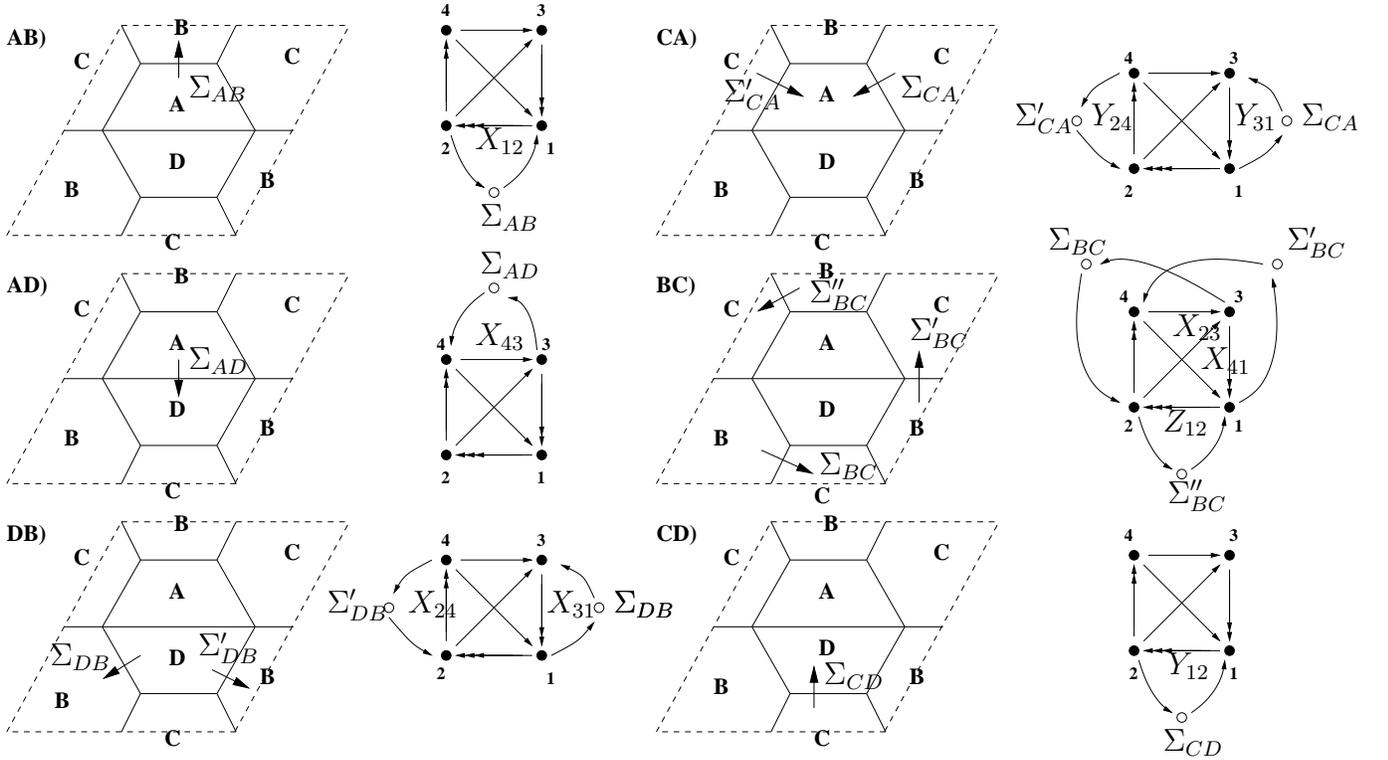}
\caption{\small
The 1-cycles and extended quivers for different kinds of D7-branes in the 
$dP_1$ theory. In the quiver we have indicated the 33 bi-fundamental 
which couples to the corresponding 37 and 73 states.}
\label{dp1d7all}
\end{figure}

Now we can construct the D7-branes associated with the different 
bi-fundamentals, and obtain the 3-7 spectrum by computing the 
intersection numbers. The different D7-branes, and the resulting extended 
quivers are shown in Figure \ref{dp1d7all}. It is a straightforward 
exercise to write down the explicit interaction terms in the presence of 
these objects (they correspond to oriented triangles in the extended quivers). 
In Figure \ref{dp1d7all} we have indicated the 33 
bifundamental which couples to the 37 and 73 states for each choice of 
D7-brane.

\begin{figure}[!htp]
\begin{center}
\epsfxsize=12cm
\hspace*{0in}\vspace*{.2in}
\epsffile{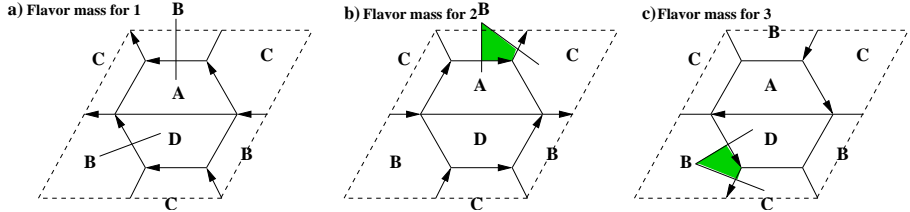}
\caption{\small
D3 and D7-branes in the $dP_1$ theory studied in Section \ref{flavdp1}. 
For each D3-brane gauge factor we have shown the D7-branes leading to the 
corresponding flavors.} 
\label{flavormass}
\end{center}
\end{figure}

In the main text we consider introducing a set of D7-branes corresponding 
to $\Sigma_{AB}$, $\Sigma_{BC}$ and $\Sigma_{DB}$. In order to determine 
the possible mass terms for the D3-brane flavors, we need to obtain the 
coupling of 77' states to states in the 37 and 37' sectors, for 
the different choices of D3-brane gauge factors. In Figure 
\ref{flavormass} we show 
the 1-cycles for the gauge factors 1, 2 and 3 (namely those with extra 
flavors), along with the D7-brane 1-cycles intersecting them. Figures (b) 
and (c) lead to disk diagrams of the kind in Figure \ref{dp0d7supo}, making 
manifest the existence of interactions leading to flavor masses. 
Figure (a) does not contain disks, but is the analog of Figure 
\ref{nodisk}. Namely there is a $37-77'-7'3$ coupling leading to masses 
for flavors of gauge factor 1. This can be shown by requiring consistency 
with the existence of such coupling in the $dP_0$ theory, upon higgssing 
the $dP_1$ theory.

\bibliographystyle{JHEP}

\end{document}